%% file: main_v2.tex
\documentclass[twocolumn,floatfix,superscriptaddress,longbibliography,nofootinbib,prx]{revtex4-2}

\input{preamble}

\begin{document}

%\preprint{APS/123-QED}

\title{Functional mesoscale organization of complex networks}

\author{Arsham Ghavasieh}
\altaffiliation{These authors contributed equally to this work}
%\email[~]{aghavasieh@fbk.eu}
\affiliation{Luddy School of Informatics, Computing, and Engineering, Indiana University, Bloomington, Indiana, 47408, USA}

\author{Satyaki Sikdar}
%\email[~]{satyaki.sikdar@gmail.com}%
\altaffiliation{These authors contributed equally to this work}
\affiliation{Department of Computer Science, Loyola University Chicago, Chicago, Illinois 60626, USA}
\affiliation{Luddy School of Informatics, Computing, and Engineering, Indiana University, Bloomington, Indiana, 47408, USA}

\author{Manlio De Domenico}
\email[Corresponding author:~]{manlio.dedomenico@unipd.it}
%\altaffiliation{Corresponding authors}
\affiliation{Department of Physics and Astronomy ``Galileo Galilei", University of Padua, Via F. Marzolo 8, 315126 Padova, Italy}
\affiliation{Padua Center for Network Medicine, University of Padua, Via F. Marzolo 8, 315126 Padova, Italy}
\affiliation{Istituto Nazionale di Fisica Nucleare, Sez., 35131 Padova, Italy}

\author{Santo Fortunato}
\email[Corresponding author:~]{santo@iu.edu}%
\affiliation{Luddy School of Informatics, Computing, and Engineering, Indiana University, Bloomington, Indiana, 47408, USA}
% \affiliation{Indiana University Network Science Institute (IUNI), Bloomington, Indiana 47408, USA}

\date{\today}% It is always \today, today,

\begin{abstract}
%Complex biological and artificial systems are pervasively characterized by modules and hierarchies, whose units exchange information to perform specific functions. 
The network density matrix (NDM) framework, enabling an information-theoretic and multiscale treatment of network flow, has been gaining momentum over the last decade. Benefiting from the counterparts of physical functions such as free energy and entropy, NDM's applications range from estimating how nodes influence network flows across scales 
%the centrality of nodes at the local level 
to explaining the emergence of structural and functional order.  Here, we introduce a generalized notion of the network internal energy $E_\tau$, where $\tau$ denotes a temporal hyperparameter allowing for multi-resolution analysis, showing how it measures the leakage of dynamical correlations from arbitrary partitions, where the minimally leaky subsystems have minimal $E_\tau$. Moreover, we analytically demonstrate that $E_\tau$ reduces to the well-known modularity function at the smallest temporal scale $\tau = 0$. We investigate this peculiar resemblance by comparing the communities minimizing $E_\tau$, with those detected by widely used methods like multiscale modularity and Markov stability. Our work provides a detailed analytical and computational picture of network generalized internal energy, and explores its effectiveness in detecting communities in synthetic and empirical networks within a unifying framework.
\end{abstract}

\maketitle

%\tableofcontents

%\begin{tcolorbox}
%\textbf{Popular summary.} \todo{popular summary}
%\end{tcolorbox}

%%%%%%%%%%%%%%%%%%%%%%%%%%%
%%%%%%%%%%%%%%%%%%%%%%%%%%%
%%%%%%%%%%%%%%%%%%%%%%%%%%%

\section{Introduction}

Complex networks consist of intricate relations among units, exhibiting interesting similarities with the entanglement of a quantum system~\cite{de2016spectral,de2019quantum_networks,nokkala2024complex}. The latter  
%A quantum system 
is often characterized by incomplete information about its state, with our best description corresponding to the probability of finding it in one state out of an orthogonal set of alternatives. In practice, their state is mixed, and an effective way to describe it is by means of a density matrix that encodes the associated statistical ensemble~\cite{Fano_1957,mcweeny1960some}. Their equilibrium thermodynamics can be faithfully described in terms of a partition function with its corresponding Von Neumann entropy~\cite{jaynes1957informationII}.

Interestingly, density matrices have proved relevant in capturing the properties of classical complex systems with interconnected structures, where some dynamical process unfolds on top of the underlying network. Early attempts to characterize such network states have proposed a density matrix based on a normalized version of the graph Laplacian matrix~\cite{braunstein2006laplacian,Severini_density_1}. This operator usually appears in the governing dynamics of many processes of physical interest, such as diffusion~\cite{Masuda_random_walk} and synchronization~\cite{arenas2008synchronization}. For instance, it has been used to describe a suitably normalized Hamiltonian operator of a non-relativistic quantum particle interacting with the quantum gravitational field within the framework of Loop Quantum Gravity~\cite{rovelli2010single}. However, interpreting the state and its entropy as genuine physical properties was difficult.

Therefore, physically meaningful density matrices that mirror the equilibrium-like mixed states have been introduced~\cite{de2016spectral,de2019quantum_networks}. The corresponding ensemble -- which allows for a multiscale analysis by means of a temporal parameter that characterizes the time scale of signaling between the system's units -- can be deduced from a statistical field theory~\cite{ghavasieh2020statistical}, and its generalization~\cite{ghavasieh2022generalized}. It has been applied to optimize transport properties~\cite{ghavasieh2020enhancing} in empirical multilayer systems~\cite{de2013mathematical}, as well as to characterize the robustness of social, biological, and transportation networks~\cite{ghavasieh_Structural_robustness,ghavasieh2021multiscale}, uncovering the core which controls information processing in a network~\cite{villegas2022paths}, to introduce a renormalization group technique based on diffusion distances~\cite{Gabrielli_RG_density, Zhang2025} and to explain the emergence of topological properties across systems~\cite{Ghavasieh2024NetForm}.

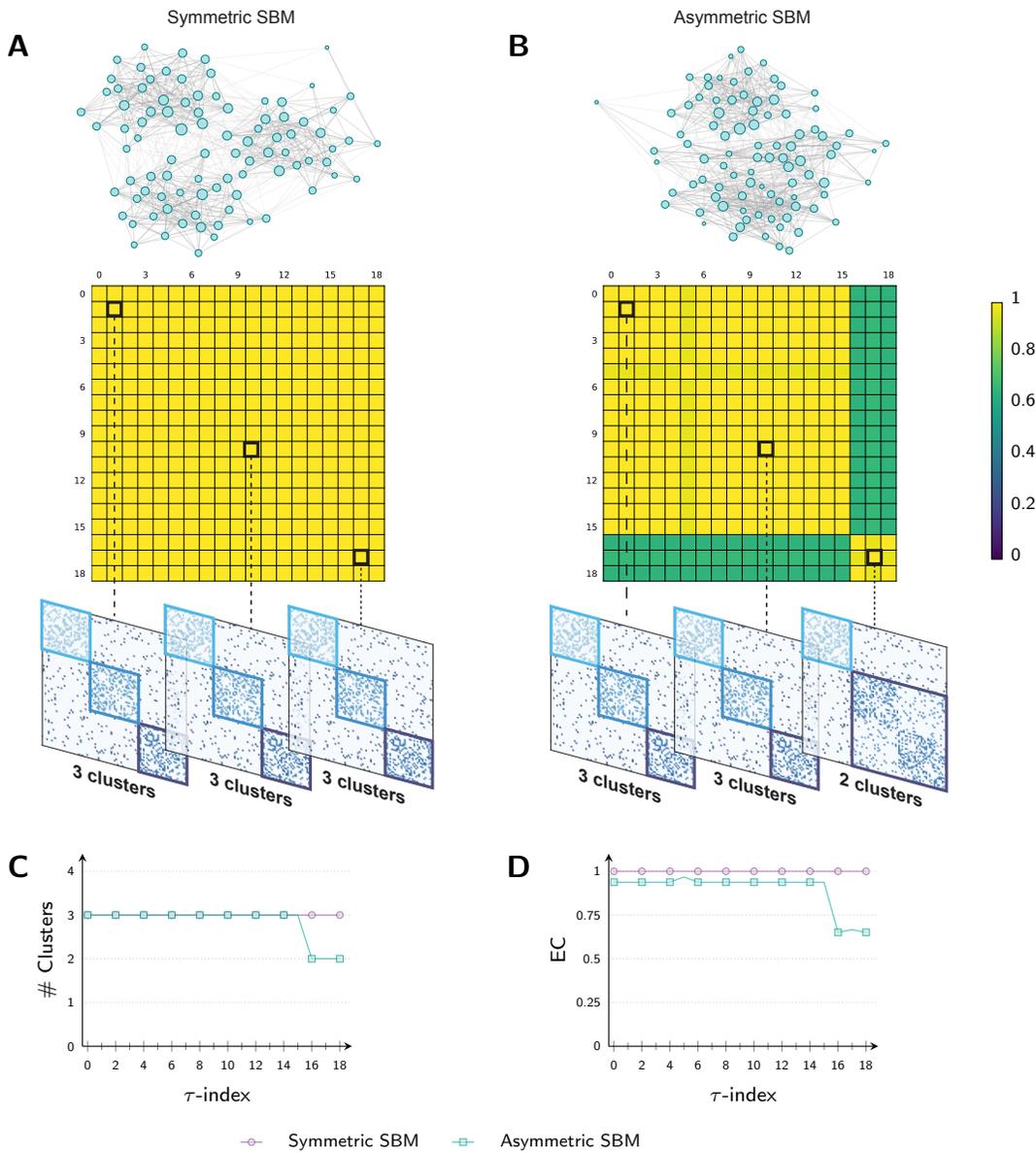
\begin{figure*}[htb]
    \input{new-figs/fig1-teaser}
    \caption{\textbf{Schematic representation of minimum internal energy clusters.} Partition similarity heatmaps show the EC similarity of pairs of partitions obtained at different temporal scales ($\tau$s), for (A) symmetric SBM with $p_{in}=0.35$ and $p_{out}=0.05$ and (B) the asymmetric SBM with $p_{in}=0.35$, $p_{out}=0.05$ except for two blocks that are connected with $p_{asymm}=p_{out}+0.05$. Colored adjacency matrices show the partitions at three different scales ($\tau$), indicating small-, middle-, and long-range communications. (C) The number of clusters, and (D) the EC similarity of partitions with the structural ground truth, as a function of $\tau$, are also reported.
    } \label{fig:Fig1}
\end{figure*}

The success of the network density matrix in linking the dynamical processes and macroscopic thermodynamic-like properties of networks, naturally raises a question: how do the thermodynamics of a network relate to its community structure?

Empirical networks often comprise groups of nodes, called communities or clusters, that strongly interact within themselves. Generally, these interactions can be short- to long-range, happening through small to large-length link sequences. Some techniques have been developed to detect these clusters at different topological scales, based on generative models and Bayesian inference~\cite{peixoto2014hierarchical}, while other methods are based on efficient heuristics that, however, lack explicit generative mechanisms. For instance, modularity optimization~\cite{newman2006modularity} 
compares networks with their randomized degree-preserving null-models, and finds clusters with a significantly high number of internal links.
%is highly successful at small scales (paths of %length $1$),  
In contrast, the multiscale modularity manually multiplies the null-model term by an arbitrary coefficient that weakens or strengthens it, and its optimization leads to solutions with smaller or larger clusters~\cite{reichardt06statistical}. More recently, the Markov stability framework~\cite{lambiotte_markov} exploited dynamical processes on top of networks to uncover communities at different scales. It starts with the steady-state distribution of a dynamical process and perturbs it to find clusters that significantly deviate from the steady state. For example, in diffusion dynamics, the steady state is uniform, with all $N$ nodes having $1/N$ share of the total concentration. The flow then starts in the proximity of the perturbed nodes, spreading to distant parts of the network until it returns to the steady state for large temporal scales $\tau \rightarrow \infty$. 

In this article, we explore the properties of the network internal energy $E_\tau$ derived from the density matrix of various networks with continuous and discrete dynamics. We analytically show that minimizing $E_\tau$ directly corresponds to minimizing the leakage rate of dynamical correlations from arbitrary groups of nodes. Furthermore, we show that at the smallest temporal scale $E_{\tau=0}$ and for diffusive dynamics, the internal energy minimization equals modularity maximization. This suggests a link between the mesoscale organization of networks and their thermodynamic-like functions. Therefore, we use a Louvain-like local agglomeration algorithm for $E_\tau$ minimization, to detect mesoscale communities in a range of synthetic and empirical systems.
We compare our findings with those found via multiscale modularity optimization method and the Markov stability framework.

\section{Generalized internal energy}

To model the dynamics of a concentration vector $\psi_\tau$ at time $\tau$, we use linear--- or linearized--- dynamical processes with control operator $\H$ and time evolution operator $\G$ in continuous 
\begin{align}\label{eq:continuous_dynamics}
    \partial_\tau\psi_\tau = -\H \psi_\tau, \qquad
    \G = e^{-\tau \H},
\end{align}
and discrete form:
\begin{align}\label{eq:discrete_dynamics}
    \psi_{\tau+1} = \H \psi_\tau, \qquad
    \G = \H^{\tau},
\end{align}
where $\psi_{\tau, i}$ encodes signal amplitude on top of node $i$ at time $\tau$. Depending on the choice of $\H$, equations can describe a range of propagation processes, like random-walks, diffusion, graph walks and consensus dynamics on top of networks (See Tab.~\ref{tab:control_operators}). The $ij$-th element of $\G$ encodes the signal amplitude received by node $i$ from node $j$, after $\tau$ time steps.

%% new table - Satyaki
\bgroup
\def\arraystretch{1.5}%  1 is the default, change whatever you need
\begin{table}[tb]
    \caption{Control operator $\H$ for a number of dynamical processes. $\A$: adjacency matrix, $\D$: degree diagonal matrix.\label{tab:control_operators}}
    \begin{tabular}{@{}ll l r@{}}
        \toprule 
        \multicolumn{2}{c}{\textbf{Dynamic}} & \phantom{abc}  & \multicolumn{1}{c}{\textbf{Operator} $\H$} \\
        \colrule 
        \multirow{2}{*}{Random Walk \phantom{}} & Continuous&& $\I - \A \D^{-1}$ \\
        & Discrete && $\A \D^{-1}$ \\[2pt]

        Diffusion & Continuous && $\D - \A$ \\[2pt]   
         Consensus & Continuous && $\I - \D^{-1}\A $\\[2pt]  
        \botrule
    \end{tabular}
\end{table}
\egroup 
%%%

Assuming that all nodes can get perturbed and send signals with equal probabilities, the statistical propagator can be obtained as $\operator{U}_\tau = \G \G^{\dag}$ with diagonal elements encoding the signal energy--- i.e., squared magnitude of the signal--- on top of nodes and off-diagonal element encoding the two-point correlations of nodes.
Note that the definition of correlation here follows the one used in quantum mechanics, that is the multiplication of two signal amplitudes--- while it can be reduced to the zero lag covariance in specific cases. Therefore, the 2 point correlation depends on their magnitudes and alignment, indicating how often and strongly they receive signals from the shared sources~\cite{ghavasieh2022generalized}.

The network density matrix is defined as $\boldsymbol{\rho}_\tau = \frac{\mathbf{U}_\tau}{Z_\tau}$,
with partition function encoding the total signal energy $Z_\tau = \tr{\operator{U}_\tau}$~\cite{ghavasieh2022generalized}. Moreover, the internal energy gives the fractional leakage rate of the total signal energy:
\begin{align}
    E_\tau = -\partial_\tau \log{Z_\tau} = -\frac{\partial_\tau Z_\tau}{Z_\tau} =- \frac{\tr{\partial_\tau \operator{U}_\tau}}{Z_\tau}.
\end{align}

Note that the diagonal elements of $\mathbf{U}_\tau$ are self-correlations or, in other words, signal energy. Therefore, the partition function gives the total signal energy and the internal energy measures the fractional leakage of signal energy. In the following, we generalize the formulation of internal energy and define the generalized internal energy that can, in addition to self-correlations, work with node-node correlations.

Let $\operator{\gamma}$ be a membership matrix with $N$ columns for the $N$ nodes of the network and $M$ rows for the $M$ labels given to the nodes. $\operator{\gamma}_{m,n}$ is $1$ only if node $n$ has label $m$, and $0$ otherwise. Note that the membership matrix must include all the nodes, and the groups do not overlap--- i.e., a node has one and only one label. The generalized internal energy reads 
\begin{align}\label{eq:partition_energy}
    E_{\tau}^{(\gamma)}= \frac{-\partial_\tau }{Z_\tau}\tr{\operator{\gamma} \operator{U}_\tau\operator{\gamma}^{\dag}} = - \frac{\tr{\operator{\gamma}(\partial_\tau \operator{U}_\tau)\operator{\gamma}^{\dag}}}{Z_\tau}.
\end{align}

Note that Eq.~~\ref{eq:partition_energy} reduces to internal energy if $\operator{\gamma}$ is an $N$-by-$N$ identity matrix.

The node-node correlations--- or, in other words, the off-diagonal elements of $\mathbf{U}_\tau$---, can leak from partition $\gamma$, according to Eq~\ref{eq:partition_energy}. In general, these correlations can be positive or negative. Note that the correlations encoded in NDMs are not based on typical measures like Pearson's. Rather, they resemble the coherence in quantum mechanics--- i.e., the multiplication of signal amplitudes
at two different points of the configuration space, translated as the space of nodes. Since the networks and dynamical processes we are concerned with in this article have positive links weights and values, our correlations are going to be non-negative. Of course, this condition can change by introducing negative edges or dynamical processes such as oscillators with amplitudes that can go negative.

Also, note that, given the exponential form of the solution for $\mathcal{G}_\tau$ (See Eq.~\ref{eq:continuous_dynamics}), is straightforward to show that $-\partial_\tau \operator{U}_\tau = \H \operator{U}_\tau + \operator{U}_\tau \H^\dag$ in the case of continuous dynamics. 

\section{Identifying minimum energy partitions}

In the following, we exploit the network internal energy to build a quality function, whose minimization by variation of $\operator{\gamma}$, gives minimally leaky clusters of nodes. 

To assess the significance of correlation leakage, we compare leakage at instance $\tau$ with the leakage at the same instance but through randomized pathways with control operator $\H^{\prime}$. 

If system's control operator $\H$ is built upon the network adjacency matrix $\A$, the noisy control operator $\H^{\prime}$ leads the flow through a randomized adjacency matrix $\A^{\prime}$. Note that we typically use the configuration model~\cite{molloy1995critical} to obtain the randomized pathway, except where we explicitly mention otherwise. As a result, $\A^{\prime}$ becomes a shuffled version of $\A$, but with the same row and column sums, \ie, node degrees. 

In case of continuous dynamics, the noisy control operator leads to a rate $\partial_{\tau}^{\prime}\operator{U}_\tau=\H^{\prime} \operator{U}_\tau - \operator{U}_\tau \H^{\prime \dag}$, indicated by a prime on top of the derivative.

The difference between leakage rate through network and noisy pathways provides a quality function for network community detection, given by 
\begin{align}\label{eq:quality_function}
    \Delta E_{\tau}^{(\gamma)} = \frac{\tr{\operator{\gamma}\operator{R} \operator{\gamma}^{\dag}}}{Z_\tau},
\end{align}

with $\operator{R}=-\partial_\tau \operator{U}_\tau+\partial_{\tau}^{\prime} \operator{U}_\tau$ encoding the difference between the system's leakage and that of its null model. If we linearize the difference we get $\operator{R}\approx\operator{R}^{(0)} - \tau \operator{R}^{(1)} $, where

\begin{align}\label{eq:linearized_R}
    \operator{R}^{(0)}=& \H + \H^{\dag} - \H^{\prime} - \H^{\prime \dag} \\ 
    \operator{R}^{(1)}=& \H^2+(\H^\dag)^2+2\H \H^\dag  \nonumber \\
    & -\H^\prime \H - \H^\prime \H^\dag-\H\H^{\prime\dag}-\H^\dag\H^{\prime\dag}
\end{align}

describes analytically the small-time scale behavior of the function (see Appendix~\ref{app:R_matrices}). For instance, in case of diffusion dynamics, we recover the negative of the modularity matrix at the smallest temporal scale ($\tau = 0$). Since our method is based on minimization and the modularity approach is based on maximization, this demonstrates that the groups of minimum general internal energy, having minimum correlation leakage, are those found by the modularity maximization method~\cite{newman2006modularity,Fortunato2022}. Similarly, for random walks, we obtain a degree-corrected version of modularity maximization at the smallest scale ($\tau=0$). Of course, at larger temporal scales, generalized internal energy minimization diverges from modularity, as it explores longer-range pathways. 

Establishing a close relation with modularity, a widely adopted method, is reassuring. Moreover, it links the generalized internal energy to established methods like Markov Stability~\cite{lambiotte_markov}, who also reduces to modularity at certain temporal scales.

However, at the same time, recovering modularity at the smallest temporal scale $\tau = 0$ indicates that internal energy minimization, if treated as a community detection technique, would suffer from the shortcomings of modularity maximization, especially in the detection of small-sized clusters below the resolution limit~\cite{fortunato2007resolution}.

Fortunately, various dynamical processes are allowed by the framework that have the potential to break this barrier. To explore a simple one, we add inertia to the random walkers. The inertia $x\geq 0$ determines the probability that a random walker or an infinitesimal quanta of a certain concentration remains static--- and excluded from the dynamical rule--- at any instance. In the more intuitive case of random walk, the ``stay" and ``move" probabilities read $p_{\text{stay}}=\frac{x}{x+1}$ and $p_{\text{move}}=\frac{1}{x+1}$, respectively. Note that $x=0$ recovers the dynamical process without inertia. Since in diffusion and random walk the flow from each node is proportional to degree, adding inertia is reduced to a transformation of the adjacency matrix $\operator{A}\rightarrow \operator{A} + x \operator{D}$, before calculating the control operator $\operator{H}$ (See Table~\ref{tab:control_operators}).

An interesting consequence of this transformation, that is nothing more than adding self-loops, is its different effect on different dynamics. Note that the addition of self-loops equally affects the diagonal elements of $\mathbf{A}$ and $\mathbf{D}$. Therefore, $\mathbf{D} - \mathbf{A}$, that is the control operator for diffusion dynamics (See Table~\ref{tab:control_operators}), remains invariant under the transformation. In contrast, the addition of self-loops directly affects the control operator of random walks $\mathbf{I} - \mathbf{A} \mathbf{D}^{-1}$, slowing it down. 

Despite the fact that the diffusion dynamics would not be affected by the transformation, the randomized control operator associated with it will be affected, as the added self-loops are randomly distributed between pairs of unconnected nodes. Inevitably, this speeds up the flow through the randomized pathways compared to the system, making the null model less strict and allowing for finer clusters. The same effect can be observed for random walks that actually slow down with the transformation.

For a complete derivation of the $\mathbf{R}$ matrix from different dynamical processes including diffusion, continuous and discrete random walks, and continuous and discrete graph walks, see Appendix~\ref{app:R_matrices}. Also, we exploit an algorithm to find clusters that minimize the $\mathbf{R}$ matrix that is explained in Appendix~\ref{app:algorithm}. Finally, while the framework is flexible to deal with different dynamical processes, in our numerical analysis we use continuous random walks with inertia. In fact, the random walk version of the Markov stability framework~\cite{lambiotte_markov} provides an ideal benchmark, since it also recovers multiscale modularity and is extensively explored. The choice of continuous random walks in our framework enables direct comparison with Markov stability framework and cross-validation, in terms of consistency, with it.

With the formalism in place, we now shift our focus on evaluating its effectiveness to unravel functional mesoscale organization in networks. 
We start with a handful of modestly sized toy networks before proceeding to the widely-used LFR benchmark~\cite{lancichinetti2008benchmark} for synthetic networks. 
Finally, we conclude the section with a thorough analysis on the real-world network of the global flights where we lack ground-truth cluster labels.
The source code of our method can be found on GitHub\footnote{\url{https://github.com/satyakisikdar/correlation-graph-clustering}}. 

\section{Clusters in synthetic and empirical networks}

% \subsection{Toy networks}
\textit{Toy networks. }
We consider three small graphs of size $N=90$: an \ER graph with edge probability $p= \tfrac{1}{9}$, and two stochastic block model (SBM) graphs. The main features of SBMs is that they have a planted group (block) structure, and that the probability of forming an edge between any two edge only depends on the group memberships of the nodes. We consider two SBMs,
\viz, \textit{symmetric} and \textit{asymmetric}, each with 3 equally-sized blocks~\cite{holland1983stochastic}.  
For the symmetric SBM we set the within-group ($p_{in}$) and across-group ($p_{out}$) edge probabilities as $0.35$ and $0.05$ respectively.  
The asymmetric SBM shares a nearly identical setup. 
We increase the across-group probabilities $p_{out}$ for two blocks by $0.05$ to increase the density of links between them.

For all three cases, we report the number of clusters and the similarity with the planted partition as a function of $\tau$ in Fig.~\ref{fig:Fig1}, panels (C) and (D). 
Partition similarity is measured using the element-centric clustering score (EC)~\cite{gates2019element}. 
EC---unlike conventional comparison methods like the widely adopted normalized mutual information---is more robust to biases introduced due to random memberships and skewed cluster sizes.
We highlight a few different partitions from the heatmaps across different time scales ($\tau$) in Fig.~\ref{fig:Fig1}.

In the case of SBMs, for example, the segregated structure and asymmetries constrain the dynamical process, giving rise to persistent clusters that coincide with the planted partition, identified through generalized internal energy minimization, which is reflected by the similarity of partitions obtained at different values of $\tau$. For instructions on indexing the $\tau$ values, see Appendix~\ref{app:algorithm}.

% \subsection{LFR \& SBM benchmarks}
\textit{LFR \& SBM Benchmarks. }
To further explore the subsystems that minimize the generalized internal energy, we analyze a set of larger synthetic networks of size $N=1000$.
We use the LFR benchmark~\cite{lancichinetti2008benchmark}, which generates networks with heterogeneous distributions of node degree and community size, with the following parameters: %  and their simplified SBM counterpart
the mixing parameter $\mu$ $\in$ $\{0.05, 0.20, 0.50, 0.60\}$; the average degree, $\langle k \rangle$ $\in$ $\{10, 15, 20\}$; the maximum degree, $k_{\max} = 50$;  community size minimum $c_{\min} = 10$ and maximum $c_{\max} = 100$; and the power-law exponents for the degree $\gamma_1 = 2$ and community size $\gamma_2 = 3$ distributions.  
The mixing parameter $\mu$ controls the strength of the community structure. 
When it is small, most connections stay inside the planted clusters, and as it grows, the clusters begin to blend until they are no longer distinguishable.
Similarly, the average degree $\langle k \rangle$ controls the sparsity of the edges, when small, there are fewer connections per node making the clusters harder to detect for a fixed $\mu$. 

Furthermore, we derive a set of SBM graphs from their LFR counterparts, having the same average degree and mixing parameters. However, they differ in two key aspects:
they all have 50 clusters of 20 nodes each, and their degree distributions are Poisson and not power laws. 
These function as an alternate, and simpler baseline than LFR.

We generate ten independent realizations for each configuration for LFR and SBM.
For each realization, we run each method ten times and record the optimal partition. 
We then report the average EC score between the optimal and the planted partitions over the ten realizations. 

\begin{figure}[htb]
    \input{new-figs/fig2-lfr-sbm-10}
    \caption{\textbf{Partition stability and quality comparison across different temporal scales for our energy minimization framework with random-walk (RW) dynamics.} 
    (A) Partition similarity heatmaps across pairs of $\tau$ values for LFR (top) and SBM (bottom) networks each with $\langle k \rangle$ 10. The columns represent different values of the mixing parameter $\mu$. 
    Similarity is assessed with the EC score, with lighter cells representing high similarity.
    RW is more stable in networks with a more pronounced community structure (low $\mu$), indicated by the blocks of high partition similarity.  
    (B) Average partition quality measured by EC score  for every combination of $\mu$ and $\tau$ index for LFR (left) and SBM (right) respectively. 
    Different values of $\mu$ are indicated with different marker styles and colors. 95\% confidence intervals around the means are shaded. 
    RW extracts the ground truth for $\mu = 0.05$ for $\tau$ index $\le$ 12. Partition quality usually saturates for a middling value of $\tau$ and then tapers off. 
    RW performs better on SBM networks compared to LFR networks potentially due to the uniformity of community sizes in SBM networks. 
    }
    \label{fig:Fig2}
\end{figure}
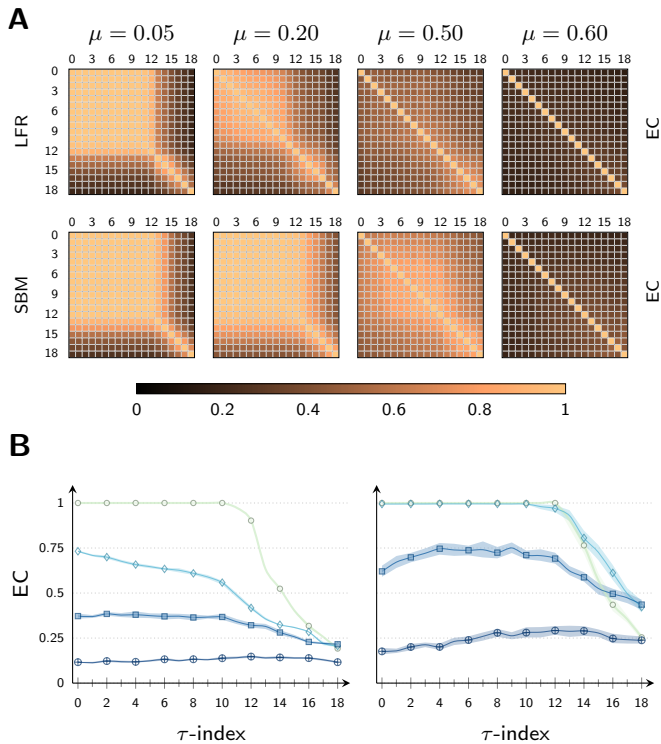

As in Fig.~\ref{fig:Fig1}, heatmaps in Fig.~\ref{fig:Fig2} indicate the similarity of partitions we obtain at different temporal scales $\tau$, indicating more partition stability and correlations with the planted partition for smaller $\mu$. 
Also, as expected, low $\mu$ leads to highly defined planted partitions reflected in the generalized energy distribution leading to higher EC scores. 
Especially, it is worth noting that in case of the LFR networks with larger $\mu$, due to the heterogeneous distribution of planted community sizes and their connectivities, the correlation between planted partition and the partition minimizing generalized energy is not strong. %(See Fig.~\ref{fig:Fig2}). 

\begin{figure*}
    \centering
    \input{new-figs/inertia-heatmaps}
    \caption{
    \textbf{Effect of inertia on performance.}
    Top row: The highest EC score obtained by RW for a given combination of $\mu$ and inertia  across $\tau$ on LFR networks with $\langle k \rangle$ = 10, 15, and 20 respectively.
    The first row shows the baseline performance, that of RW \textbf{without} inertia. Higher values (yellow) indicate higher similarity with the ground truth partition. 
    Bottom row: The relative improvement of EC score obtained by using inertia over the baseline. 
    The cells are gray where the baseline perfectly recovered the ground truth, when $\mu = 0.05$.
    We observe a marked performance improvement across the board by adding inertia. The gains saturate quickly with increasing values of inertia, implying diminishing returns. 
    }
    \label{fig:inertia}
\end{figure*}
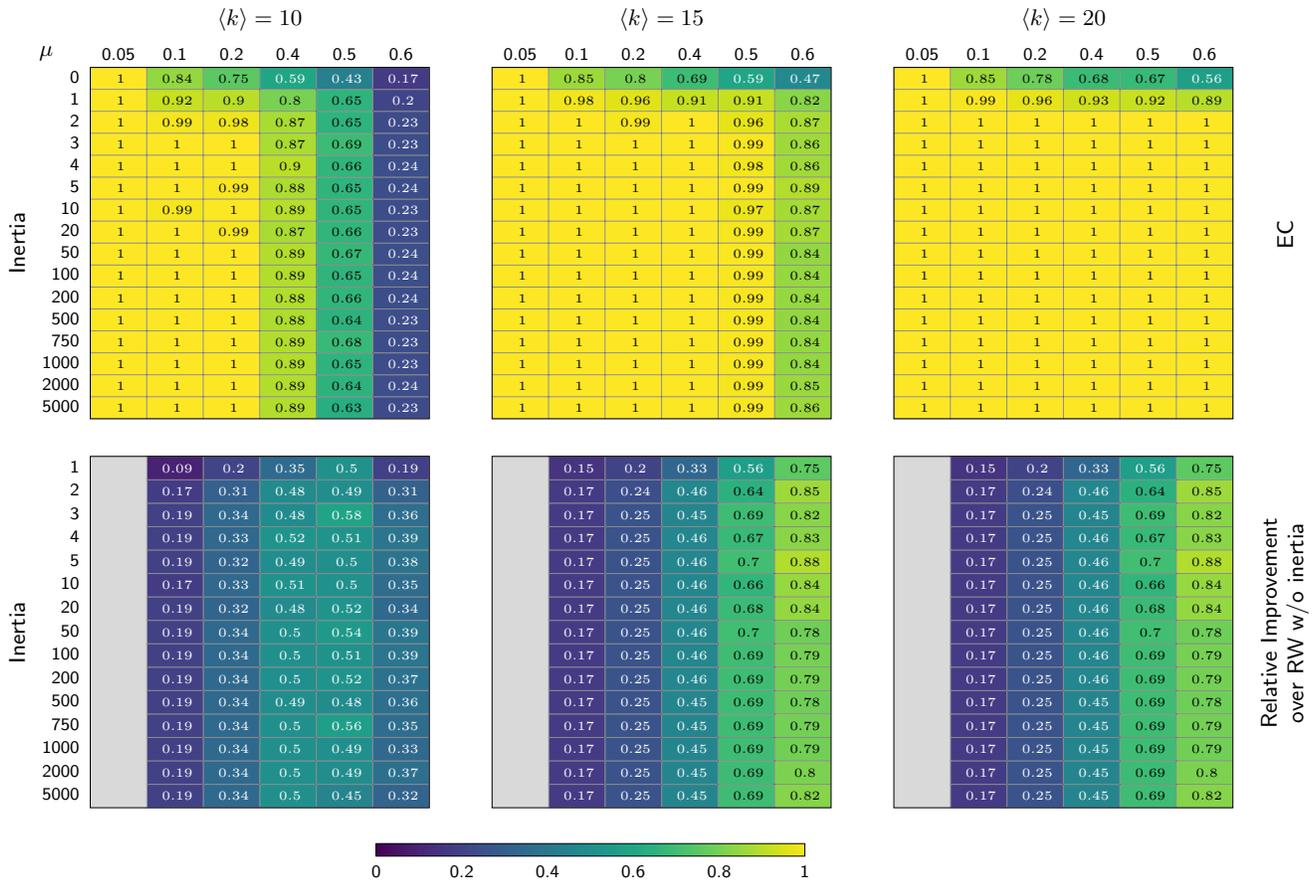

As explained in the previous section, the addition of inertia is expected to improve the visibility of planted clusters to the dynamical process. Indeed, as shown in Fig.~\ref{fig:inertia}, slowing down the dynamical process via inertia dramatically increases the correlation between minimal energy partitions and the planted ones as shown in the top panel.
The heatmaps in the bottom panel quantifies this improvement, by measuring the gain in EC score obtained as a result of adding inertia relative to the performance without it. 
Indeed, inertia provides a much-needed boost providing an EC score gain of 40-50\%.  

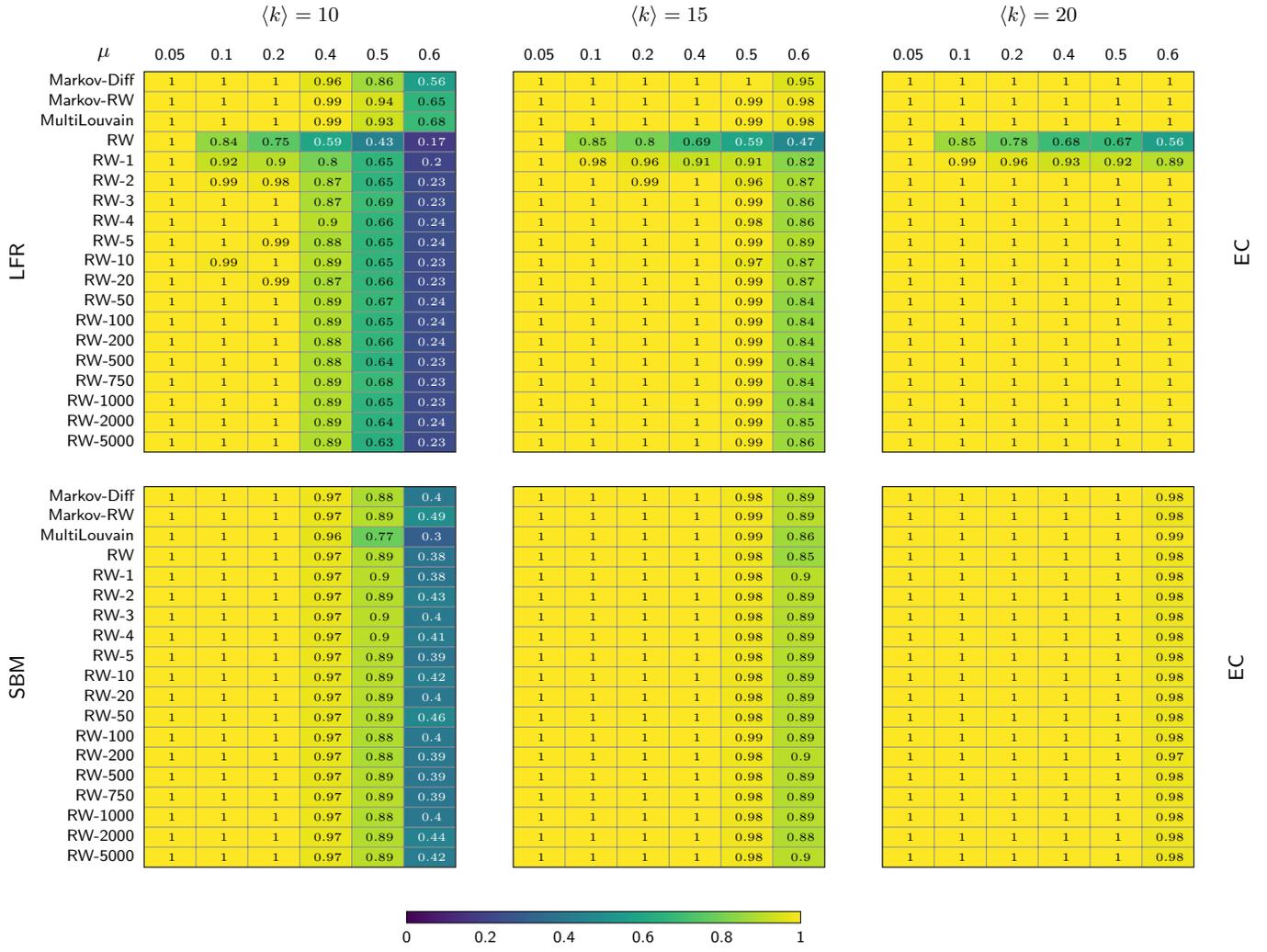
\begin{figure*}
   \centering
   \input{new-figs/methods-comparison-heatmap}
   \caption{\textbf{Performance comparison between different methods.}
   Each method is evaluated on LFR (top row) and SBM (bottom row) networks with different combinations of $\mu$ and $\langle k \rangle$. We consider the highest EC score obtained by each method across $\tau$ with respect to the ground truth partition. 
   A perfect retrieval yields an EC score of 1. 
   The first three rows in each heatmap represent the performance of related methods: Markov-Diff, Markov-RW, and MultiLouvain.
   Different values of inertia ($i$) for RW are indicated as RW-$i$. 
   In most cases, our proposed method is able to match or even beat the performance of the other methods.
   }
   \label{fig:inertia_comparison}
\end{figure*}

Furthermore, this improvement is best captured on a ring-of-cliques network, like the one in Fig.~\ref{fig:ring_of_clique}, first introduced to test the resolution limit of modularity maximization~\cite{fortunato2007resolution}.
Our method perfectly recovers the ground truth, with each clique placed in its own cluster, when we add inertia. 
Similarly, Fig.~\ref{fig:inertia_comparison} provides a complementary analysis comparing how effective the internal energy minimization is in capturing the planted partition with Markov stability and multiscale modularity frameworks, indicating that internal energy minimization recovers the ground truth better than or as well as others, even without inertia, in the case of stochastic block models, while it requires inertia to catch up in the case of LFR networks. 
 
\textit{Flights Network. }
We further evaluate our method on a real-world network of non-stop flights constructed using the freely available OpenFlights dataset\footnote{\url{https://openflights.org/}}. 
This weighted network consists of 18,859 edges (non-stop flights) across 3,214 nodes (airports), with the edge weights reflecting the frequency of flights between a pair of airports.
In the absence of ground truth, we adapt the Silhouette score~\cite{rousseeuw1987silhouettes} to assess the cohesiveness and separation of each clustering computed based on the Haversine distance. 
The Haversine distance uses the latitude and longitude coordinates of the airports to provide a proxy for geographical proximity between nodes. 
These measures are formally defined in Appendix~\ref{app:sil}.
Given a clustering, we compute the weighted and unweighted mean Silhouette score of nodes within each cluster to obtain a single score between -1 and +1. In the former, each cluster's mean is assigned a weight equal to the number of members.
Regardless of the weighting strategy, a higher value of the Silhouette score signifies every node being well-matched (nearer) with other nodes in its own cluster and poorly matched (farther) with nodes in neighboring clusters.

Results indicate that minimum energy partitions, with inertia, are significantly better in detecting the nodes that are geographically closer (See Fig.~\ref{fig:flight_silhouette}) for both variants of the score. Figure.~\ref{fig:map} provides a visualization of such a partition.
We find the clusters roughly reflect distinct global regions, only London Heathrow (LHR) is put in the same cluster (light green) as airports in the Middle East, Africa, and Southeast Asia.

Note that this experiment does not indicate that the internal energy minimization is inherently superior to the well-known methods like multiscale modularity and Markov stability, since the internal energy framework is benefiting from the additional parameter, namely inertia, that slows down the dynamical process. Otherwise, as previous theoretical and numerical results indicate, the performance of the minimum internal energy partition in detecting the planted clusters--- or their approximations--- is comparable with the aforementioned methods. Therefore, our work must be considered a theoretical attempt that find connections between thermodynamic-like functions of networks like internal energy and the mesoscale organization reliably detected by well-known methods.

\begin{figure}[htb]
    \centering
    \input{new-figs/cliq-ring-inertia}
    \caption{\textbf{Resolution limit of energy minimization.} 
    The highest quality partition obtained by energy minimization on a network of 30 cliques of size 5 ($K_5$) arranged in a ring. 
    It has an EC score of 0.56 with 17 clusters containing a mix of individual $K_5$'s (in blue) and pairs of $K_5$'s (orange). 
    We obtain the ground truth partition, with each $K_5$ in separate clusters, when the inertia is set to any value $>$ 0. 
    The other methods, Markov stability and multiscale modularity also recover the ground truth for suitable values of $\tau$. 
    }
    \label{fig:ring_of_clique}
\end{figure}
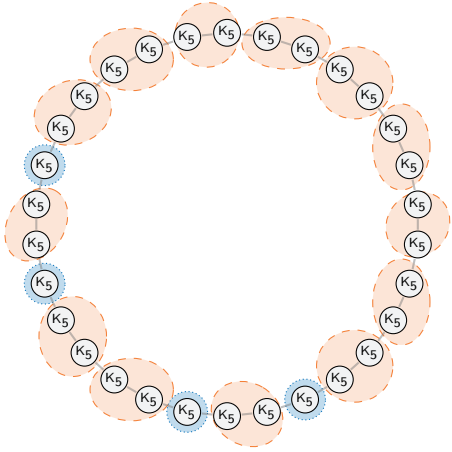

\begin{figure*}[htb!]
    \centering
    \input{new-figs/flights-world-sil-bar}
    % \caption{Flight worlds mean silhouette and mean weighted silhouette scores. Inertia v1: self-loops $\propto$ degrees. Inertia v2 are hidden, none are in either top 10 lists.}
    \caption{\textbf{Clustering performance on the flights network.} 
    % \satyaki{replace bars with a hair pin plot?}
    The top 16 methods with the highest mean unweighted (left, blue) and mean weighted (right, pink) Silhouette Coefficient obtained across all $\tau$ on the global Flights network.
    A higher score indicates more compact and well-separated clusters based on Haversine distance. 
    Different values of inertia ($i$) are represented as RW-$i$. 
    The different variants of  find significantly \textit{better} clusters than the other methods.  
    }
\label{fig:flight_silhouette}
\end{figure*}
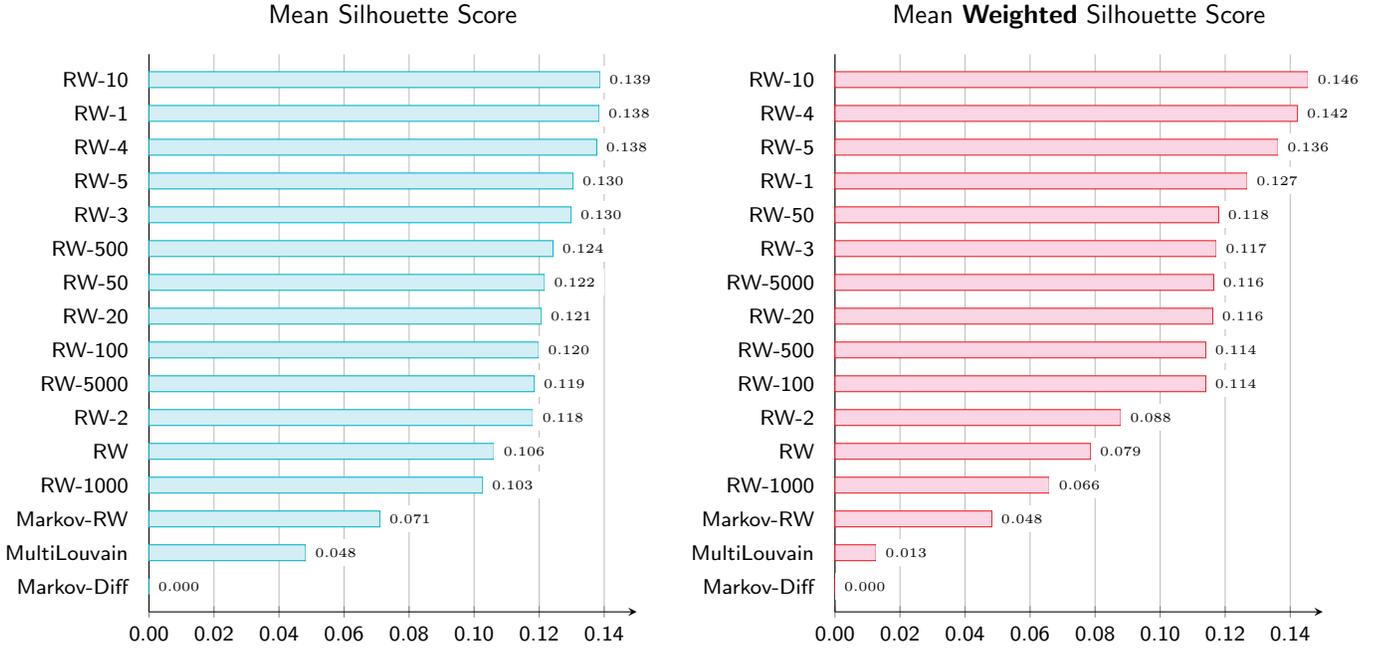

\begin{figure*}[htb!]
    \centering
    % trim: left, bottom right top
    % \fbox{\includegraphics[trim={0 9.6cm 0 7.5cm}, clip, width=0.85\textwidth]{new-figs/flight-test.pdf}}
    \includegraphics[width=0.9\textwidth]{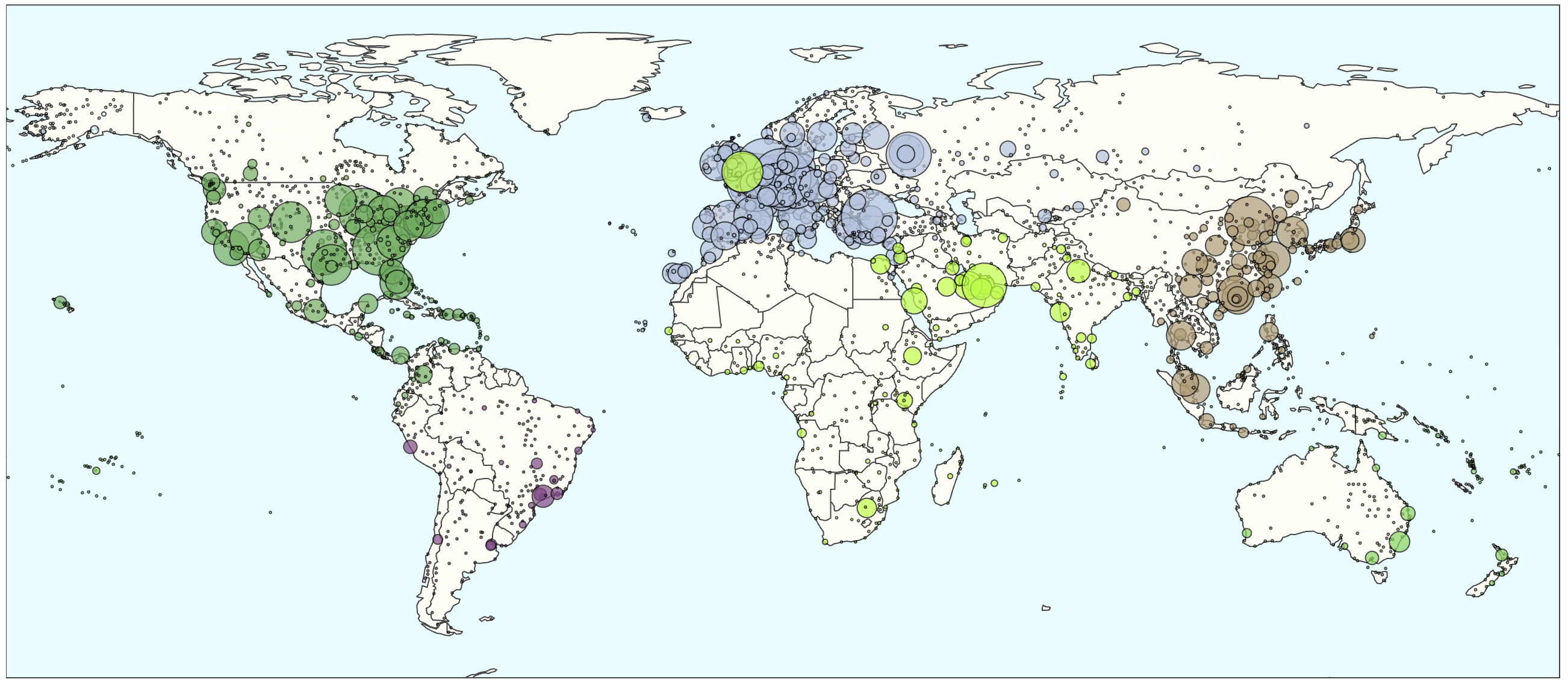}
    \caption{\textbf{Flights network clusters found by energy minimization.} 
    This clustering with 15 clusters has the highest value of mean Silhouette score (unweighted: 0.11, weighted: 0.146) on the Flights network across all methods. 
    Nodes (airports) in the same cluster have the same color, and their sizes are proportional to their degrees. Edges are not drawn for clarity. 
    We observe that clusters largely correspond to global regions spread across continents. 
    Interestingly, London Heathrow (LHR) is placed in the same cluster as the Middle East and Southeast Asia (light green), and not with other airports in Europe (blue). This is likely due to a high-volume flights between LHR and parts of Asia and the Middle East.
    }
    % \caption{RW, inertia 10 v1, tau ix: 12, best performing method overall. Mean silhouette: 0.11, weighted mean silhouette: 0.146. 15 clusters.}
    \label{fig:map}
\end{figure*}

\section{Conclusion}

In this work we studied the network generalized internal energy $E_\tau^{(\gamma)}$, a thermodynamic‐like functional coming from the density matrix formulation, showing that it quantifies how leaky the dynamical correlation are out of an arbitrary partition $\gamma$ at any temporal scale ~$\tau$.  We analytically demonstrate that, for diffusive dynamics, minimizing $E_{\tau}$ at the smallest temporal scale ($\tau = 0$) is exactly equivalent to maximizing the well-known modularity function, suggesting a link to the community detection problem. Of course, at larger temporal scales $\tau$ the correspondence breaks, allowing the method to probe the multiscale organization along paths of increasing length.  We use a Louvain–style greedy descent to vary partitions while minimizing the energy $E_\tau$, and obtain partitions that closely resemble the ones found by the well-known algorithms like Markov stability~\cite{lambiotte_markov} and multiscale modularity maximization~\cite{reichardt06statistical}.

While we focus on diffusive dynamics, the framework is general enough to be applicable to other dynamical systems, including non-linear ones, that can be explored in the future by using the same formalism proposed in~\cite{ghavasieh2022generalized}.

Previous studies have covered the thermodynamic-like functions such as the Von Neumann entropy and free energy~\cite{ghavasieh2022generalized,Ghavasieh2024NetForm}, with applications to microscopic and macroscopic features. Our work provides an understanding of network internal energy and explores its correspondence with networks' mesoscale organization. We show how minimizing internal energy recovers well-established methods of community detection, like multiscale modularity, as special cases. This tie, at the same time, explains the limitations of internal energy minimization when dealing with finer sized clusters. As an example of the versatility of the framework, we show how modifying the dynamical process by adding inertia can improve the performance. Our work opens the doors for follow-ups exploiting our framework combined with other dynamical processes, including the non-linear ones, to study dynamical communities for the analysis of large empirical systems. Also, it paves the way for a deeper understanding of network thermodynamics, information flow and structural organization.

\begin{acknowledgements}
    A. G. and S. F. acknowledge the support of the AccelNet-MultiNet program, a project of the National Science Foundation (Award \#1927425 and \#1927418). 
\end{acknowledgements}
    
\newpage
\appendix

\section{R matrices}\label{app:R_matrices}
Multiscale modularity (MM), Markov stability (MS) and generalized internal energy are based in optimization of quality functions that can be written in form of a matrix $\operator{R}$, sandwiched between the membership matrix $\operator{\gamma}$ and its transpose as $\tr{\operator{\gamma}\operator{R} \operator{\gamma}^{\dag}}$. While MM and MS try to maximize this mathematical expression by variation of the membership matrix, generalized internal energy method minimizes it, searching for minimum partition internal energy.

\subsection{Multiscale modularity}

For MM, the R matrix is given by
\begin{align}\label{eq:MM_quality}
    \operator{R}_{\text{MM}}=\operator{A}-\operator{A}^{\prime}/\tau,
\end{align}
where $\tau$ is a tunable parameter determining the scale, $\operator{A}$ is the adjacency matrix and $\operator{A}^{\prime}$ is the degree-preserved randomized null model, also called the configuration model, with $\A^{\prime}_{ij} = k_i k_j/2m$, where $m$ is the total number of links and $k_i$ is the degree of node $i$, in the original network.

\subsection{Markov stability}

Whereas, for Markov stability, this matrix is given by 
\begin{align}\label{eq:MS_quality}
    \operator{R}_{\text{MS}}=\operator{\Pi}\G -\pi^{\dagger} \pi,
\end{align}
where $\pi$ is the stationary-state distribution of the process that multiplied by its conjugate becomes the MS null-model. Also, steady-state can be represented like $\operator{\Pi} = diag(\pi)$, as a diagonal matrix. $\mathcal{G}_\tau = e^{-\tau \operator{H}}$ is the time-evolution operator for continuous dynamics, with its discrete counterpart $\mathcal{G}_\tau = \operator{H}^{\tau}$.
The choice of dynamical processes affects both null-model $\pi^{\dagger}\pi$ and the perturbation matrix $\operator{\Pi}\G$. 

Here we mainly work use MS with continuous random walk dynamics, where the null model become $\pi^{\dagger}\pi=\operator{A}^{\prime}/m$, with $\pi_i=k_i/m$ and the control parameter reads $\H=\I - \A \D^{-1}$, with $\D$ being the degree diagonal matrix $D_{ij} = \delta_{ij}k_i$. In this case, it has been shown that Markov stability recovers MM formalism at small scales $\tau\ll 1$.

We also report results from MS with continuous diffusion dynamics, where the null model is a matrix whose all elements are $1/N^{2}$ for a network of $N$ nodes, with $\pi_i=1/N$ and 
the control operator reads $\H=\D - \A$.

%For continuous dynamics, $\operator{R}_{\text{PIE}}$ that we write as $\operator{R}$ for brevity, follows
%\begin{align}\label{eq:PIE_R_Continuous}
 %   \operator{R} &= \H \operator{U}_\tau + \operator{U}_\tau \H^\dag - \H^{\prime} \operator{U}_\tau - \operator{U}_\tau \H^{\prime \dag} \\
  %  \operator{U}_\tau &= \G \G^{\dag}, \nonumber
%\end{align}
%where $\H^{\prime}$ is the noisy control operator from the randomized pathways of configuration model $\operator{A}^{\prime}$. It is also possible to replace the configuration model with an Erdos-Renyi null-model, whose all element read $m/N(N-1)$. 

\subsection{Continuous generalized internal energy}

In case of continuous dynamics, the R matrix is given by
\begin{eqnarray}
    \operator{R}&=&-\partial_\tau \operator{U}_\tau+\partial_{\tau}^{\prime} \operator{U}_\tau \\
    \operator{U}_\tau&=&\mathcal{G}_\tau\mathcal{G}_\tau^\dag. \\
    -\partial_\tau \operator{U}_\tau &=&\H \operator{U}_\tau + \operator{U}_\tau \H^\dag
\end{eqnarray}

To further explore its behavior, we linearize the time-evolution operator $\mathcal{G}_\tau=e^{-\tau \H}\approx \I-\tau \H$, we can expand the propagator $\operator{U}_\tau=\mathcal{G}_\tau\mathcal{G}_\tau^\dag$ and its derivative $-\partial_\tau \operator{U}_\tau =\H \operator{U}_\tau + \operator{U}_\tau \H^\dag$. 

For the the propagator at small $\tau$, we obtain 
\begin{align*}
    \operator{U}_\tau &\approx (1 - \tau \H+\mathcal{O}(\tau^2))(1 - \tau \H^\dag+\mathcal{O}(\tau^2))  \\
    &\approx 1 - \tau \left(\H + \H^\dag\right) + \mathcal{O}(\tau^2)   
\end{align*}

Directly, the derivative reads 
{
\small 
\begin{align}
    -\partial_\tau \operator{U}_\tau =&\ \H + \H^\dag - \tau \left(\H^2+(\H^\dag)^2+2\H \H^\dag\right) + \mathcal{O}(\tau^2)
    \label{eq:continuous_leakage}
\end{align}
}

with a simpler representation for hermitian control operators $\H=\H^\dag$, reading
\begin{equation}\label{eq:continuous_hermitian_leakage}
    -\partial_\tau \operator{U}_\tau = 2\H - 
    4\tau \H^2 + \mathcal{O}(\tau^2).
\end{equation}

Similarly, the correlation leakage rate through the noisy control operator $\H^{\prime}$ reads

\begin{align}
    \begin{split} \label{eq:continuous_leakage_null}
        -\partial_\tau^\prime \operator{U}_\tau ={}&  \H^\prime+\H^{\prime\dag} \\
        &-\tau \left(\H^\prime \H + \H^\prime \H^\dag+\H\H^{\prime\dag}+\H^\dag\H^{\prime\dag}\right)  \\
        & + \mathcal{O}(\tau^2),
    \end{split}
    \intertext{which if both control operators are hermitian, simplifies into}
    -\partial_\tau^\prime \operator{U}_\tau =&\ 2\H^\prime - 2\tau \left(\H^\prime \H + \H \H^\prime\right) + \mathcal{O}(\tau^2), \label{eq:continuous_hermitian_leakage_null}
\end{align}

Therefore, the linearization of $\operator{R}=-\partial_\tau \operator{U}_\tau+\partial_{\tau}^{\prime} \operator{U}_\tau$ follows $\operator{R}\approx \operator{R}^{(0)}-\tau \operator{R}^{(1)}$, with

\begin{align}\label{eq:linearized_R}
    \operator{R}^{(0)}=& \H + \H^{\dag} - \H^{\prime} - \H^{\prime \dag} \\ 
    \operator{R}^{(1)}=& \H^2+(\H^\dag)^2+2\H \H^\dag  \nonumber \\
    & -\H^\prime \H - \H^\prime \H^\dag-\H\H^{\prime\dag}-\H^\dag\H^{\prime\dag}.
\end{align}
valid for small values of $\tau$.

To ensure the validity, the propagation scale $\tau$ should not exceed a certain value. To estimate it, Assume the eigenvalues of the control parameter $\H$ are ${\epsilon_{\ell}}$, all non-negative, and sorted in ascending order such that $\epsilon_{\ell}<\epsilon_{\ell^\prime}$ if $\ell < \ell^\prime$. A linearization of the time-evolution operator is given by $\mathcal{G}_\tau=e^{-\tau \H}\approx \I-\tau \H$ requiring the $\ell$-th eigenvalue to follow $\tau \epsilon_\ell \ll 1 \rightarrow e^{-\tau \epsilon_{\ell}}\approx 1 - \tau \epsilon_\ell$. Therefore, the largest eigenvalue $\epsilon_{N}$ determines the validity of linearization, imposing the constraint $\tau < 1/\epsilon_{N}$. We define the middle temporal scale, $\tau_{mid}=1/\epsilon_{N}$, determining the point at which the dynamics deviates from the linear approximation.

\subsubsection{Continuous Diffusion}\label{app:diff_C}

In this case, the continuous dynamics is governed $\H=\D - \A$, leading to a correlation leakage of 

\begin{equation}\label{eq:diffusion_leakage}
    -\partial_\tau \operator{U}_\tau \approx 2\D - 2\A - 4\tau \left(\D^2+\A^2+\A \D+\D \A \right).
\end{equation}
Note that in optimization of $\gamma$, the diagonal matrices in the correlation leakage rate does not contribute. Therefore, for the purpose of minimizing the leakage rate by changing the membership matrix, one can discard the diagonal matrices and write Eq~\ref{eq:diffusion_leakage} as $2\A-4\tau(\A^2+\A \D+\D \A)$.

The null model control operator here is $\H^\prime=\D^\prime - \A^\prime$, where it can be shown that $\D^\prime=\D$. Therefore, 
\begin{eqnarray}
    \operator{R}^{(0)}= -2 \left( \A - \A^\prime \right),
\end{eqnarray}

whose element-wise representation is given by 
\begin{eqnarray}
    R^{(0)}_{ij}= -2 \left( A_{ij}-\frac{k_i k_j}{2m} \right),
\end{eqnarray}
showing that minimizing the negative modularity or, equally, maximizing the modularity, is recovered as the first order approximation of the metric.

Also, the correction term reads
\begin{align}
    \begin{split}
        \operator{R}^{(1)} ={}& 4\left(\D - \A\right)^2 -2\left(\D - \A^\prime\right)\left( \D - \A\right) \\ 
        &-2\left(\D - \A\right)\left(\D - \A^\prime\right) \nonumber \\
        ={}& 4\left(\D^2-\D \A - \A \D + \A^2\right) \nonumber \\
        & -2\left(\D^2-\A^\prime\D-\D \A + \A^\prime\A\right)  \nonumber \\ 
        & -2\left(\D^2-\A \D-\D \A^\prime + \A \A^\prime\right) \nonumber \\
    \end{split}
    \\
    \begin{split}
        \operator{R}^{(1)} ={}& 4 \A^2 - 2\A \A^\prime - 2\A^\prime\A +2 \A^\prime \D + 2 \D \A^\prime \\
        &  -2\D \A-2\A \D,
    \end{split}
\end{align}

whose element-wise representation is given by 
\begin{eqnarray}
    R^{(1)}_{ij}&&= 2\sum\limits_{k=1}^{N} (2 A_{ik}A_{kj} -  A_{ik}\frac{k_k k_j}{2m}\nonumber\\
    &&- \frac{k_i k_k}{2m}A_{kj}-k_i A_{ij} -A_{ij}k_{j}  ).
\end{eqnarray}

Note that the correction term is $-\tau\tr{\gamma\operator{R}^{(1)}\gamma^\dag}=-\tau\sum\limits_{i,j=1}^{N}R_{ij}^{(1)}\delta_{c(i),c(j)}$, where $\delta_{c(i),c(j)}$ is zero unless $i$ and $j$ are encoded to be in the same group by $\gamma$. 

\subsubsection{Continuous Random Walk}\label{app:RW_C}

In this case, the continuous dynamics is governed by $\H=\I - \A \D^{-1}$, where $\D$ is the degree diagonal matrix $\langle i|\D|j\rangle = \delta_{ij}k_i$, leading to a correlation leakage rate of 

\begin{eqnarray}
    -\partial_\tau \operator{U}_\tau &&\approx \I - \A \D^{-1} + \I - \D^{-1}\A \nonumber \\
    &&4\tau[ (\I - \A \D^{-1})^2 + \nonumber \\ &&(\I - \D^{-1}\A)^2 + \nonumber \\
    &&(\I - \D^{-1}\A)(\I - \A \D^{-1}) + \nonumber \\ &&(\I - \A \D^{-1}) (\I - \D^{-1}\A)] ,
\end{eqnarray}
for undirected networks $\A=\A^\dag$. The expression can be represented as

\begin{eqnarray}\label{eq:communicability_leakage}
    -\partial_\tau \operator{U}_\tau &&\approx \I - \A \D^{-1} + \I - \D^{-1}\A + \nonumber \\
    &&4\tau[ \I+\A \D^{-1}\A \D^{-1}- 2\A \D^{-1} + \nonumber \\ &&\I+\D^{-1}\A \D^{-1}\A - 2 \D^{-1}\A + \nonumber \\
    &&\I - \D^{-1}\A - \A \D^{-1} + \D^{-1}\A \A \D^{-1} + \nonumber \\ &&\I - \A \D^{-1}-\D^{-1}\A + \A \D^{-1}\D^{-1}\A ] ,
\end{eqnarray}
which can be simplified by discarding diagonal matrices
\begin{eqnarray}\label{eq:communicability_leakage}
    -\partial_\tau \operator{U}_\tau &&\approx -\A \D^{-1} -\D^{-1}\A + \nonumber \\
    4\tau[&& \A \D^{-1}\A \D^{-1}- 2\A \D^{-1} + \nonumber \\ &&\D^{-1}\A \D^{-1}\A - 2 \D^{-1}\A + \nonumber \\
    &&-\D^{-1}\A - \A \D^{-1} + \D^{-1}\A \A \D^{-1} + \nonumber \\ &&-\A \D^{-1}-\D^{-1}\A + \A \D^{-1}\D^{-1}\A ] ,
\end{eqnarray}

The null model control operator here is $\H^\prime=\I - \A^\prime\D^{-1}$, where it can be shown that $\D^\prime=\D$. Therefore, 
\begin{eqnarray}
    \operator{R}^{(0)}= -(\A - \A^\prime)\D^{-1}-\D^{-1}(\A - \A^\prime),
\end{eqnarray}
whose element-wise representation is given by 
\begin{eqnarray}
    R^{(0)}_{ij}= (A_{ij}-\frac{k_i k_j}{2m})/k_j + (A_{ij}-\frac{k_i k_j}{2m})/k_i .
\end{eqnarray}

The correction term reads
\begin{eqnarray}
    \operator{R}^{(1)}=&& (\I - \A \D^{-1})^2+(\I - \D^{-1}\A)^2\nonumber \\ &&+(\I - \A \D^{-1})(\I - \D^{-1}\A)\nonumber\\
    &&+(\I - \D^{-1}\A)(\I - \A \D^{-1}) \nonumber \\
    &&- (\I - \A^\prime\D^{-1})(\I - \A \D^{-1})\nonumber\\
    &&-(\I - \A^\prime\D^{-1})(\I - \D^{-1}\A)\nonumber\\
    &&-(\I - \A \D^{-1})(\I - \D^{-1}\A^\prime)\nonumber\\
    &&-(\I - \D^{-1}\A)(\I - \D^{-1}\A^\prime)
\end{eqnarray}
which can be expanded as
\begin{eqnarray}
    \operator{R}^{(1)}&&=\I-2\A \D^{-1}+\A \D^{-1}\A \D^{-1} \nonumber\\
    &&+\I-2\D^{-1}\A+\D^{-1}\A \D^{-1}\A\nonumber\\
    &&+\I - \A \D^{-1}-\D^{-1}\A + \A \D^{-2}\A\nonumber\\
    &&+\I - \D^{-1}\A - \A \D^{-1}+\D^{-1}\A^2\D^{-1}\nonumber\\
    &&-\I+\A^\prime\D^{-1}+\A \D^{-1}-\A^\prime\D^{-1}\A \D^{-1}\nonumber\\
    &&-\I+\A^\prime\D^{-1}+\D^{-1}\A - \A^\prime\D^{-2}\A\nonumber\\
    &&-\I+\A \D^{-1}+\D^{-1}\A^\prime - \A \D^{-2}\A^\prime\nonumber\\
    &&-\I+\D^{-1}\A+\D^{-1}\A^\prime - \D^{-1}\A \D^{-1}\A^\prime
\end{eqnarray}

resulting in 
\begin{eqnarray}
    \operator{R}^{(1)}&&=   \A \D^{-1}\A \D^{-1} + \D^{-1}\A \D^{-1}\A\nonumber\\
    &&+\A \D^{-2}\A+\D^{-1}\A^2\D^{-1} \nonumber\\
    && -\A^\prime\D^{-1}\A \D^{-1}-\A^\prime\D^{-2}\A \nonumber\\
    && -\A \D^{-2}\A^\prime - \D^{-1}\A \D^{-1}\A^\prime\nonumber\\
    &&-2(\A \D^{-1}-\A^\prime\D^{-1}+\D^{-1}\A-\D^{-1}\A^\prime),
\end{eqnarray}
and considering the previous derivation $\operator{R}^{(0)}= (\A - \A^\prime)\D^{-1}+\D^{-1}(\A - \A^\prime)$, can be written represented as

\begin{eqnarray}
    \operator{R}^{(1)}&&=   \A \D^{-1}\A \D^{-1} + \D^{-1}\A \D^{-1}\A\nonumber\\
    &&+\A \D^{-2}\A+\D^{-1}\A^2\D^{-1} \nonumber\\
    && -\A^\prime\D^{-1}\A \D^{-1}-\A^\prime\D^{-2}\A \nonumber\\
    && -\A \D^{-2}\A^\prime - \D^{-1}\A \D^{-1}\A^\prime\nonumber\\
    &&-2\operator{R}^{(0)},
\end{eqnarray}

\subsubsection{Continuous Graph Walk}\label{app:GW_C}

In this case, known as communicability in the literature, the continuous dynamics is directly governed by the adjacency matrix $\H=-\A$, leading to a correlation leakage rate of 

\begin{eqnarray}\label{eq:communicability_leakage}
    -\partial_\tau \operator{U}_\tau \approx - 2\A - 
    4\tau \A^2,
\end{eqnarray}
for undirected networks $\A=\A^\dag$.

The null model control operator here is $\H^\prime= -\A^\prime$. Therefore, 
\begin{eqnarray}
    \operator{R}^{(0)}= -2 \left( \A - \A^\prime \right)
\end{eqnarray}
whose element-wise representation is given by 
\begin{eqnarray}
    R^{(0)}_{ij}= -2 \left(A_{ij}-\frac{k_i k_j}{2m}\right),
\end{eqnarray}
showing that minimizing the negative modularity or, equally, maximizing the modularity, is recovered as the first order approximation of the metric. Such optimization is done through minimizing $\tr{\gamma\operator{R}^{(0)}\gamma^\dag}=\sum\limits_{i,j=1}^{N}R_{ij}^{(0)}\delta_{c(i),c(j)}$, where $\delta_{c(i),c(j)}$ is zero unless $i$ and $j$ are encoded to be in the same group by $\gamma$. 

Here, the correction term reads
\begin{eqnarray}
    \operator{R}^{(1)}= 4\A^2 - 2 \left( \A^\prime\A + \A \A^\prime \right),
\end{eqnarray}
whose element-wise representation is given by 
\begin{eqnarray}
    R^{(1)}_{ij}= 2\sum\limits_{k=1}^{N} \left( 2 A_{ik}A_{kj} -  A_{ik}\frac{k_k k_j}{2m}- \frac{k_i k_k}{2m}A_{kj} \right).
\end{eqnarray}

\subsection{Discrete generalized internal energy}\label{app:discrete_dyn}

In case of discrete dynamics $\mathcal{G}_\tau = \H^\tau,(\tau=0,1,2,...)$, the propagator is given by $\operator{U}_\tau=\mathcal{G}_\tau\mathcal{G}_\tau^\dag$ and its negative rate reads $\frac{\Delta \operator{U}_\tau}{\Delta \tau}]_{\Delta \tau=1}=\H \operator{U}_\tau \H^\dag-\operator{U}_\tau$.

It is straightforward to see the smallest temporal scale behavior of the statistical propagator at $\tau = 0$
    
\begin{eqnarray}\label{eq:continuous_hermitian_leakage}
    - \frac{\Delta \operator{U}_{\tau=0}}{\Delta \tau} = -\H \H^\dag + 
    \I,
\end{eqnarray}

and, consequently, the R matrix
\begin{eqnarray}
    \operator{R} = - \frac{\Delta \operator{U}_{\tau=0}}{\Delta \tau} + \frac{\Delta^\prime \operator{U}_{\tau=0}}{\Delta \tau} = -\H \H^\dag + \H^\prime \H^{\prime\dag}.
\end{eqnarray}

%\begin{figure*}[htb]
 %   \includegraphics[scale=0.3]{figures/lfr_ec_ground-truth.png}

  %  \includegraphics[scale=0.35]{figures/LFR-heatmaps}
%    \caption{First row: comparison with ground truth. Below: LFR stability heatmaps. We have 20 independent LFR networks (over 1,000 nodes) for each $\mu$ and average degree values, so we plot the mean of the 20 heatmaps. \satyaki{pick different $\mu$s?}}
%\end{figure*}

%\begin{figure*}[t]
%    \centering
%    \begin{subfigure}{0.95\textwidth}
%        \includegraphics[scale=0.25]{figures/SBM-asymm_ground-trurth.png}
 %       \caption{(left) SBM graph with 90 nodes and 3 clusters (right) Adjacency matrix where nodes are colored based on community memberships.}
 %   \end{subfigure} \hfill \\
  %  \begin{subfigure}{0.95\textwidth}
   %     \includegraphics[scale=0.28]{figures/SBM-asymm_methods.png}

    %    \caption{Clusters discovered by methods for varying $\tau$}
 %   \end{subfigure}
    %\caption{Teaser figure draft. \satyaki{Get feedback}}
%\end{figure*}

\subsubsection{Discrete Graph Walk}\label{app:GW_D}

In this case, the discrete dynamics is directly governed by the adjacency matrix $\H=\A$. If the propagation scale is set to be the smallest possible value $\tau = 0$, the correlation leakage rate becomes

\begin{eqnarray}\label{eq:communicability_leakage}
    - \frac{\Delta \operator{U}_{\tau=0}}{\Delta \tau} = -\A^2 +
    \I,
\end{eqnarray}
for undirected networks $\A=\A^\dag$.

Here, the leakage rate compared to null model can be written as
\begin{eqnarray}
    \operator{R} = -\A^2 + \A^{\prime2},
\end{eqnarray}

whose element-wise representation reads
\begin{eqnarray}
    R_{ij}= \sum\limits_{k=1}^{N} -A_{ik}A_{kj} + \frac{k_i k_{k}^2 k_j}{4m^2}.
\end{eqnarray}

\subsubsection{Discrete Random Walk}\label{app:RW_D}

In this case, the discrete dynamics is directly governed by $\H=\A \D^{-1}$, where $\D$ is the degree diagonal matrix $\langle i|\D|j\rangle = \delta_{ij}k_i$, leading to a correlation leakage rate of 

\begin{eqnarray}
    - \frac{\Delta \operator{U}_{\tau=0}}{\Delta \tau} = -\A \D^{-2}\A +
    \I,
\end{eqnarray}
for undirected networks $\A=\A^\dag$. The expression can be represented as

Here, the leakage rate compared to null model can be written as
\begin{eqnarray}
    \operator{R} = -\A \D^{-2}\A + \A^\prime\D^{-2}\A^\prime.
\end{eqnarray}

whose element-wise representation reads
\begin{eqnarray}
    R_{ij}= \sum\limits_{k=1}^{N} -\frac{A_{ik}A_{kj}}{k_{k}^2} + \frac{k_i k_j}{4m^2}.
\end{eqnarray}

\section{Algorithm}\label{app:algorithm}

Note that the partition internal energy is the contains both diagonal and off-diagonal elements of $\operator{\gamma}\operator{R}\operator{\gamma}^{\dagger}$. However, variation of the membership matrix $\operator{\gamma}$ only changes the contribution of off-diagonal elements, as long as $\operator{\gamma}$ remains a valid membership matrix including all nodes and giving each one of them one and only one label. In other words, the optimization is only concerned with the leakage rate of correlations, keeping the leakage rate of signal energy as a constant term. Also, the coefficient $1/Z_\tau$ remains the same through minimization of the quality function, having no effect on the final partitions.

We use a greedy algorithm similar to Louvain to minimize the quality function. We initially assign each node to a distinct community. We randomly select one node at each step of the algorithm and calculate the change in the quality function, in case the node leaves its present community and joins each of the other communities. The node leaves its community and joins the community that minimizes the change in the quality function, if this change is negative. If there is no community shifts for two iterations over all nodes, the algorithm stops and $\operator{\gamma}$ is found as the best partition. Similarly, we use this optimization method for multiscale modularity and Markov stability. 

In the main text, for brevity, we focus on Random Walk dynamics for generalized internal energy and Markov stability (MS). We also compare them with multiscale modularity (MM) algorithm.

Moreover, to make comparisons across method and dynamics possible, we set the maximum temporal scale to be $\tau_{max}=1/\epsilon_{2}$, where $\epsilon_{2}$ is the smallest non-zero eigenvalue of $\H$, as a control operator with non-negative spectrum. $\tau_{max}$ is also called diffusion time, being the temporal scale for convergence to steady-state and typically used to rescale propagation time. 

\section{Metrics} \label{app:sil}

\textbf{Haversine Distance. } Given two points $A = (\phi_1, \lambda _1)$ and $B = (\phi_2, \lambda_2)$ on the Earth's surface, where $\phi$ denotes the latitude and $\lambda$ the longitude in radians respectively. We define the Haversine distance between $A$ and $B$ as:
$$
2r \cdot \arcsin\left( \sqrt{ hav(\phi_2 - \phi_1) + \cos(\phi_1)\cos(\phi_2)hav(\lambda_2 - \lambda_1) } \right),
$$
where $r$ is the Earth's radius and $hav(\theta) = \sin^2\left(\frac{\theta}{2}\right)$ is the Haversine function.

\textbf{Silhouette Coefficient. } Given a clustering $C$ of $k$ clusters and a distance metric $d$, we define two scores $a(i)$ and $b(i)$ for each data point $i$ in Cluster $C_i$ as follows.
\begin{align*}
    a(i) &= \frac{1}{|C_i| - 1} \sum_{\substack{j \in C_i \\ i \ne j}} d(i, j)    \\
    b(i) &= \min_{j \ne i} \frac{1}{|C_j|} \sum_{j \in C_j} d(i, j),
\end{align*}
where $a(i)$ is the mean distance between $i$ and all other data points in its cluster, and $b(i)$ is the mean distance between $i$ and the points in the nearest cluster $C_j \ne C_i$.
In other words, $a(i)$ measures how well $i$ is assigned to its cluster, the smaller the better, and $b(i)$ the next best fit cluster for point $i$, the higher the better. 

Finally, we define the Silhouette value of a data point $i$ as:
$$
s(i) = 
\begin{cases}
    \frac{b(i) - a(i)}{\max\{a(i), b(i) \}}, & \text{ if } |C_i| > 1 \\
    0, & \text{ otherwise}
\end{cases}
$$
The mean $s(i)$ over all points of a cluster is a measure of how tightly grouped all the points in the cluster are. 
Similarly, the mean $s(i)$ over all points of the entire dataset is a measure of how appropriately the data have been clustered, with higher values indicating better overall cluster quality. 
We can also consider different weighting strategies for computing the overall Silhouette score for a clustering. In the main text, we use the number of cluster members $|C_i|$ as the weight of cluster $C_i$ to compute the weighted mean Silhouette score.

\bibliography{biblio}

\end{document}

%% file: new-figs/fig1-teaser.tex
\pgfplotsset{colormap/viridis high res}

\pgfplotstableread{
name method orig_clusters tau_ix num_clusters modularity EC
SBM-symm RW 3 0 3 0.4186 1
SBM-symm RW 3 1 3 0.4186 1
SBM-symm RW 3 2 3 0.4186 1
SBM-symm RW 3 3 3 0.4186 1
SBM-symm RW 3 4 3 0.4186 1
SBM-symm RW 3 5 3 0.4186 1
SBM-symm RW 3 6 3 0.4186 1
SBM-symm RW 3 7 3 0.4186 1
SBM-symm RW 3 8 3 0.4186 1
SBM-symm RW 3 9 3 0.4186 1
SBM-symm RW 3 10 3 0.4186 1
SBM-symm RW 3 11 3 0.4186 1
SBM-symm RW 3 12 3 0.4186 1
SBM-symm RW 3 13 3 0.4186 1
SBM-symm RW 3 14 3 0.4186 1
SBM-symm RW 3 15 3 0.4186 1
SBM-symm RW 3 16 3 0.4186 1
SBM-symm RW 3 17 3 0.4186 1
SBM-symm RW 3 18 3 0.4186 1
}{\SBMsymmStats}

\pgfplotstableread{
name method orig_clusters tau_ix num_clusters modularity EC
SBM-asymm RW 3 0 3 0.3758 0.9376
SBM-asymm RW 3 1 3 0.3758 0.9376
SBM-asymm RW 3 2 3 0.3758 0.9376
SBM-asymm RW 3 3 3 0.3758 0.9376
SBM-asymm RW 3 4 3 0.3758 0.9376
SBM-asymm RW 3 5 3 0.3746 0.9678
SBM-asymm RW 3 6 3 0.3758 0.9376
SBM-asymm RW 3 7 3 0.3758 0.9376
SBM-asymm RW 3 8 3 0.3758 0.9376
SBM-asymm RW 3 9 3 0.3758 0.9376
SBM-asymm RW 3 10 3 0.3758 0.9376
SBM-asymm RW 3 11 3 0.3758 0.9376
SBM-asymm RW 3 12 3 0.3758 0.9376
SBM-asymm RW 3 13 3 0.3758 0.9376
SBM-asymm RW 3 14 3 0.3758 0.9376
SBM-asymm RW 3 15 3 0.3758 0.9376
SBM-asymm RW 3 16 2 0.2994 0.6508
SBM-asymm RW 3 17 2 0.2986 0.6667
SBM-asymm RW 3 18 2 0.2994 0.6508
}{\SBMAsymmStats}

\pgfplotstableread{
name method orig_clusters tau_ix num_clusters modularity EC
ER-90-0.111 RW 7 0 6 0.267 0.3852
ER-90-0.111 RW 7 1 6 0.2625 0.3992
ER-90-0.111 RW 7 2 6 0.2617 0.4586
ER-90-0.111 RW 7 3 6 0.2649 0.4223
ER-90-0.111 RW 7 4 6 0.2605 0.4012
ER-90-0.111 RW 7 5 7 0.2719 0.3697
ER-90-0.111 RW 7 6 6 0.2663 0.4293
ER-90-0.111 RW 7 7 6 0.2617 0.3847
ER-90-0.111 RW 7 8 6 0.2654 0.3838
ER-90-0.111 RW 7 9 6 0.2654 0.4301
ER-90-0.111 RW 7 10 6 0.2631 0.4349
ER-90-0.111 RW 7 11 5 0.2626 0.4154
ER-90-0.111 RW 7 12 6 0.2684 0.4351
ER-90-0.111 RW 7 13 5 0.2615 0.4233
ER-90-0.111 RW 7 14 6 0.2609 0.4221
ER-90-0.111 RW 7 15 5 0.2637 0.3962
ER-90-0.111 RW 7 16 5 0.2586 0.398
ER-90-0.111 RW 7 17 4 0.2552 0.3513
ER-90-0.111 RW 7 18 4 0.2541 0.359
}{\ERStats}

\tikzstyle{sbm_sym} = [
    smooth, 
    Purple,
    mark=*, 
    % mark options={fill=Pastel-H},%, point meta=explicit symbolic, text=black,
    mark options={thin, draw=Purple!80!black, fill=.!20, scale=0.65},
    draw opacity=0.7,
    fill opacity=0.7,
    mark repeat = 2, 
]

\tikzstyle{sbm_asym} = [
    JungleGreen,
    mark=square*, 
    % mark options={fill=Pastel-H},%, point meta=explicit symbolic, text=black,
    mark options={thin, draw=JungleGreen!80!black, fill=.!20, scale=0.65},
    draw opacity=0.7,
    fill opacity=0.7,
    mark repeat = 2, 
]

\tikzstyle{er} = [
    Apricot,
    mark=diamond*, 
    % mark options={fill=Pastel-H},%, point meta=explicit symbolic, text=black,
    mark options={thin, draw=Apricot!40!red, scale=0.75},
    draw opacity=0.7,
    fill opacity=0.7,
    mark repeat = 2, 
]

% \fbox
{
\begin{tikzpicture}
    \begin{scope}  % this is for the heatmaps and the top 2/3rds of the figure 
        \node at (0, 0) {\includegraphics[width=0.9\textwidth, trim={0 20pt 0 0}]{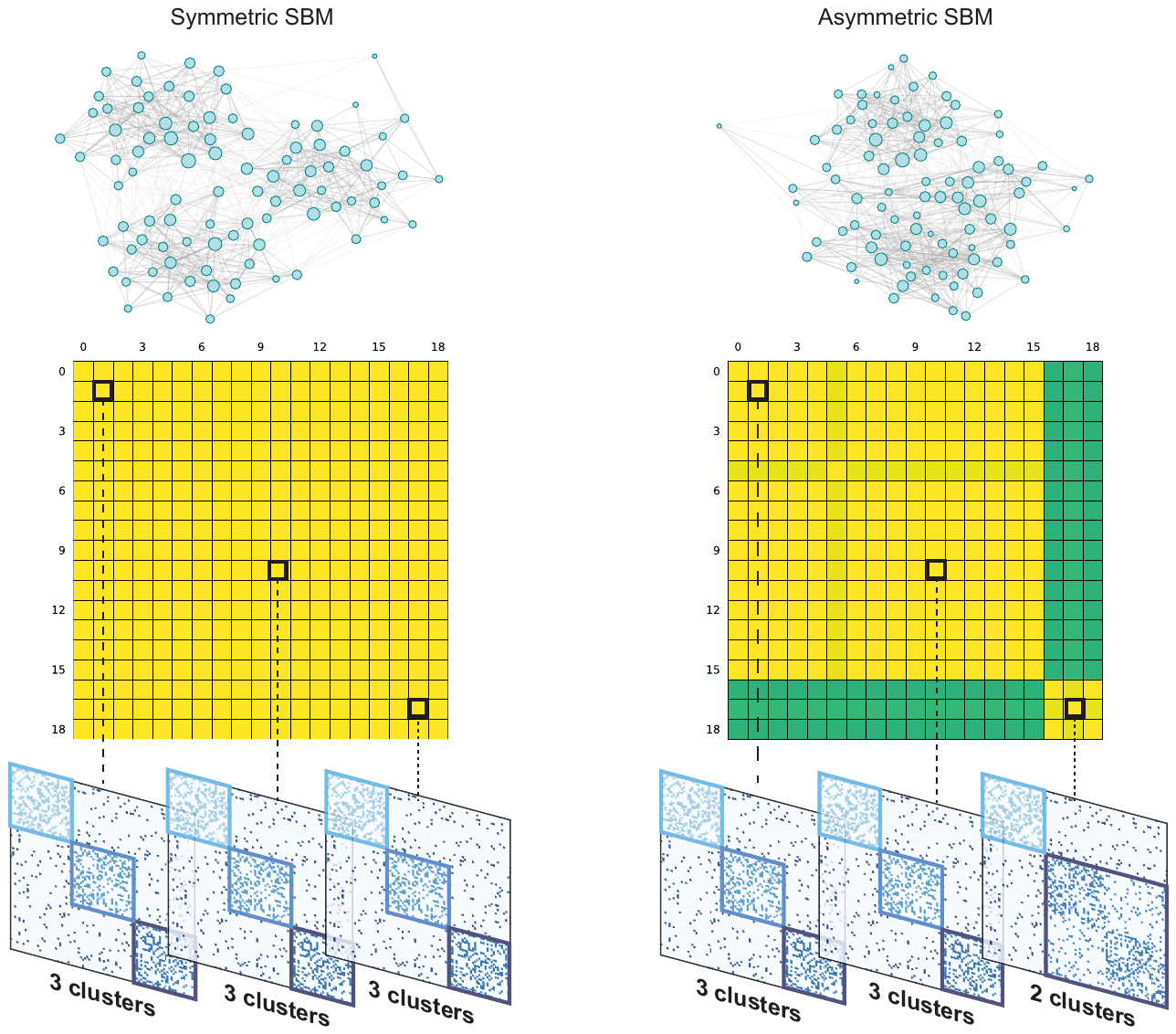}};
    \end{scope}
    
    \begin{scope}[yshift=-24em, xshift=-17.5em]
    % add the pgfplots for #clusters, modularity and EC score
    \begin{groupplot}[
        group style={
    		group name=my plots,
    		rows=1,
            columns=2,
            % x descriptions at=edge bottom,
            % y descriptions at=edge left,
            xticklabels at=all,
            xlabels at=edge bottom,
            yticklabels at=all,
            ylabels at=edge left,
    		vertical sep=20pt,
            horizontal sep=10em,
        },
        height=120pt,
        width=150pt,
        ymajorgrids,
	   grid style={draw=gray!50, draw opacity=0.75,densely dotted},
        xtick distance=2,
        minor x tick num=1,
        xtick align=center , 
        xtick style={font=\tiny\sffamily},
	ytick style={draw=none, font=\tiny\sffamily},
        xticklabel style={
            font=\tiny\sffamily,
        },
        yticklabel style={
            xshift=0.25em,
            font=\tiny\sffamily,
            /pgf/number format/.cd,
            fixed,
            % fixed zerofill,
            precision=2,
            /tikz/.cd
    	},
        xtick pos=lower,
        ytick pos=left,
        title style={
            yshift=-0.5em, 
            align=center, 
            font=\footnotesize\sffamily, 
        },
        xmin=0, xmax=18.5,
        ymin=0, ymax=1,
        % axis background/.style={fill=gray!2},
        axis x line=bottom,
        axis y line=left,
        enlarge x limits=0.02,
        enlarge y limits={value=0.1, upper},
        xlabel={$\tau$-index},
        ylabel style={
            font=\footnotesize\sffamily,
            yshift=-0.25em,
        },
        xlabel style={
            font=\footnotesize\sffamily
        },
        every node near coord/.append style={font=\tiny},
        nodes near coords align={below},
        yticklabel={\pgfmathparse{\tick}\pgfmathprintnumber[assume math mode=true]{\pgfmathresult}},
        legend columns=-1,
        legend pos=outer north east,
        legend style={
            font=\scriptsize,
            shift={(-5em, -10.5em)},
            column sep=8pt,
        },
        xtick={},
    ]

    \nextgroupplot[
        ymin=0, ymax=4, ylabel={\# Clusters}, ytick distance=1,
        yticklabel={\pgfmathparse{\tick}\pgfmathprintnumber[assume math mode=true]{\pgfmathresult}},
    ]   %%% LFR 1k RW 
        \addplot[sbm_sym] table[x=tau_ix, y=num_clusters] {\SBMsymmStats}; \addlegendentry{Symmetric SBM~~}
        
        \addplot[sbm_asym] table[x=tau_ix, y=num_clusters] {\SBMAsymmStats}; \addlegendentry{Asymmetric SBM~~}

        % \addplot[er] table[x=tau_ix, y=num_clusters] {\ERStats}; \addlegendentry{Erd\"{o}s R\'{e}nyi}
        
    % \nextgroupplot[ytick distance=0.10, ymax=0.5, ylabel={Modularity},]   %%% Modularity
    %     \addplot[sbm_sym] table[x=tau_ix, y=modularity] {\SBMsymmStats}; 
        
    %     \addplot[sbm_asym] table[x=tau_ix, y=modularity] {\SBMAsymmStats};

    %     \addplot[er] table[x=tau_ix, y=modularity] {\ERStats}; 

    \nextgroupplot[ytick distance=0.25, ylabel={EC},]   %%% LFR 1k RW 
        \addplot[sbm_sym] table[x=tau_ix, y=EC] {\SBMsymmStats}; 
        
        \addplot[sbm_asym] table[x=tau_ix, y=EC] {\SBMAsymmStats};

        % \addplot[er] table[x=tau_ix, y=EC] {\ERStats}; 
    \end{groupplot}
    \end{scope}

\begin{scope}[shift={(6, 1.75)}, scale=1]
    \pgfplotscolorbardrawstandalone[
        colorbar,
        point meta min=0, point meta max=1,
        colormap name=viridis high res,
        colorbar style={
            width=4pt,
            height=100pt,
            scale=1,
            ytick style={
                draw=none,
                /pgf/number format/.cd,
                fixed,
                % fixed zerofill,
                precision=2,
                /tikz/.cd
        	},
            yticklabel={\pgfmathparse{\tick}\pgfmathprintnumber[assume math mode=true]{\pgfmathresult}},
            yticklabel style={font=\scriptsize\sffamily, yshift=0.2em},
            yticklabel pos=right,
            scaled ticks=false,
        },
    ]
\end{scope}

\node[font=\bfseries\sffamily\large] at (-20em, 15em) {A};
\node[font=\bfseries\sffamily\large] at (-0.5em, 15em) {B}; 
% \node[font=\bfseries\sffamily\large] at (11em, 15em) {C}; 

\node[font=\bfseries\sffamily\large] at (-20em, -17em) {C};
\node[font=\bfseries\sffamily\large] at (-0.5em, -17em) {D}; 
% \node[font=\bfseries\sffamily\large] at (12.5em, -17.5em) {F}; 

\end{tikzpicture}
}

%% file: new-figs/fig2-lfr-sbm-10.tex
\input{data/CL_data}
\input{data/heatmap_data}
\pgfplotsset{colormap/GnBu}
\pgfplotsset{colormap/copper}

\begin{tikzpicture}[scale=0.95, transform shape]
    \begin{groupplot}[
        group style={
            rows=2,
            columns=4,
                group name=mygroup,
    	    vertical sep=15pt,
                horizontal sep=7.5pt,
                % x descriptions at=edge bottom,
                % xlabels at=edge top,
                xticklabels at=all,
            y descriptions at=edge left,
        },
        % width=0.50\textwidth,
        height=50pt, width=50pt,
        scale only axis=true,
        axis on top,
        colormap name=copper,
        xmin=-0.5, xmax=18.5,
        ymin=-0.5, ymax=18.5,
        % xtick distance=1, ytick distance=1,
        % xtick distance=2, ytick distance=2,
        xtick distance=3, ytick distance=3,
        typeset ticklabels with strut,
        nodes near coords align={anchor=center},
        nodes near coords style={font=\tiny, /pgf/number format/.cd,fixed,fixed zerofill,precision=1},
        % nodes near coords style={font=\scriptsize},
        % yticklabel={\pgfmathparse{\tick*100}\pgfmathprintnumber[assume math mode=true]{\pgfmathresult}\%},
        title style={align=center, yshift=0em, font=\small\sffamily},
        enlargelimits=false,
        ylabel style={font=\scriptsize\sffamily, yshift=-0em},
        xlabel style={font=\scriptsize\sffamily, yshift=-0em},
        % colorbar
        tick align=outside,
        xticklabel pos=top,
        xtick style={draw=none},
        ytick style={draw=none},
        xticklabel style={draw=none, font=\tiny\sffamily, yshift=-0.7em},
        yticklabel style={draw=none, font=\tiny\sffamily, xshift=0.35em},
        % axis background/.style={fill=gray!90},
        axis line style={draw=black},
        point meta min=0,
        point meta max=1,
    ]
    %%%%%%%%% LFR-1k, avg_k = 10
    \nextgroupplot[
        title={$\mu = 0.05$},
        ylabel={LFR},
    ]
        \addplot[
            matrix plot,  % not matrix plot* 
            draw=black, line width=0.01mm, draw opacity=0.10, 
            fill opacity=1,
            colorbar source,
            mesh/cols=19,
            point meta=explicit,
        ]
        table[x=tau_ix_x, y=tau_ix_y, meta=mean_EC] {\LFROneKkTENmuFIVEmethRW};

    % %%%%%%%%%%%% mu = 10
    % \nextgroupplot[
    %     title={$\mu = 0.10$},
    % ]
    %     \addplot[
    %         matrix plot,  % not matrix plot* 
    %         draw=black, line width=0.01mm, draw opacity=0.10,
    %         % draw=black, line width=0.005mm,
    %         fill opacity=1,
    %         colorbar source,
    %         mesh/cols=19,
    %         point meta=explicit,
    %     ]
    %     table[x=tau_ix_x, y=tau_ix_y, meta=mean_EC] {\LFROneKkTENmuTENmethRW};

    %%%%%%%%%%%% mu = 20
    \nextgroupplot[
        title={$\mu = 0.20$},
    ]
        \addplot[
            matrix plot,  % not matrix plot* 
            draw=black, line width=0.01mm, draw opacity=0.10,
            fill opacity=1,
            colorbar source,
            mesh/cols=19,
            point meta=explicit,
        ]
        table[x=tau_ix_x, y=tau_ix_y, meta=mean_EC] {\LFROneKkTENmuTWENTYmethRW};

    % %%%%%%%%%%%% mu = 40
    % \nextgroupplot[
    %     title={$\mu = 0.40$},
    % ]
    %     \addplot[
    %         matrix plot,  % not matrix plot* 
    %         draw=black, line width=0.01mm, draw opacity=0.10,
    %         % draw=black, line width=0.005mm,
    %         fill opacity=1,
    %         colorbar source,
    %         mesh/cols=19,
    %         point meta=explicit,
    %     ]
    %     table[x=tau_ix_x, y=tau_ix_y, meta=mean_EC] {\LFROneKkTENmuFORTYmethRW};
    
    %%%%%%%%%%%%
    \nextgroupplot[
        title={$\mu = 0.50$},
    ]
        \addplot[
            matrix plot,  % not matrix plot* 
            draw=black, line width=0.01mm, draw opacity=0.10,
            fill opacity=1,
            colorbar source,
            mesh/cols=19,
            point meta=explicit,
        ]
        table[x=tau_ix_x, y=tau_ix_y, meta=mean_EC] {\LFROneKkTENmuFIFTYmethRW};

    %%%%%%%%%%%%
    \nextgroupplot[
        title={$\mu = 0.60$},
        ylabel={EC},
        ylabel style={at={(current axis.east)}, anchor=center, yshift=-1.25em},
    ]
        \addplot[
            matrix plot,  % not matrix plot* 
            draw=black, line width=0.01mm, draw opacity=0.10,
            fill opacity=1,
            colorbar source,
            mesh/cols=19,
            point meta=explicit,
        ]
        table[x=tau_ix_x, y=tau_ix_y, meta=mean_EC] {\LFROneKkTENmuSIXTYmethRW};
    
    %%%%%%% second row 
    %%%%%%%%%%%%%%% SBM avg k = 5
    \nextgroupplot[
        ylabel={SBM},
    ]
        \addplot[
            matrix plot,  % not matrix plot* 
            draw=black, line width=0.01mm, draw opacity=0.10, 
            fill opacity=1,
            colorbar source,
            mesh/cols=19,
            point meta=explicit,
        ]
        table[x=tau_ix_x, y=tau_ix_y, meta=mean_EC] {\SBMOneKkTENmuFIVEmethRW};

    % %%%%%%%%%%%% mu = 10
    % \nextgroupplot[
    %     title={$\mu = 0.10$},
    % ]
    %     \addplot[
    %         matrix plot,  % not matrix plot* 
    %         draw=black, line width=0.01mm, draw opacity=0.10,
    %         % draw=black, line width=0.005mm,
    %         fill opacity=1,
    %         colorbar source,
    %         mesh/cols=19,
    %         point meta=explicit,
    %     ]
    %     table[x=tau_ix_x, y=tau_ix_y, meta=mean_EC] {\SBMOneKkTENmuTENmethRW};

    %%%%%%%%%%%% mu = 20
    \nextgroupplot[
    ]
        \addplot[
            matrix plot,  % not matrix plot* 
            draw=black, line width=0.01mm, draw opacity=0.10,
            fill opacity=1,
            colorbar source,
            mesh/cols=19,
            point meta=explicit,
        ]
        table[x=tau_ix_x, y=tau_ix_y, meta=mean_EC] {\SBMOneKkTENmuTWENTYmethRW};

    % %%%%%%%%%%%% mu = 40
    % \nextgroupplot[
    %     title={$\mu = 0.40$},
    % ]
    %     \addplot[
    %         matrix plot,  % not matrix plot* 
    %         draw=black, line width=0.01mm, draw opacity=0.10,
    %         % draw=black, line width=0.005mm,
    %         fill opacity=1,
    %         colorbar source,
    %         mesh/cols=19,
    %         point meta=explicit,
    %     ]
    %     table[x=tau_ix_x, y=tau_ix_y, meta=mean_EC] {\SBMOneKkTENmuFORTYmethRW};
    
    %%%%%%%%%%%%
    \nextgroupplot[
    ]
        \addplot[
            matrix plot,  % not matrix plot* 
            draw=black, line width=0.01mm, draw opacity=0.10,
            fill opacity=1,
            colorbar source,
            mesh/cols=19,
            point meta=explicit,
        ]
        table[x=tau_ix_x, y=tau_ix_y, meta=mean_EC] {\SBMOneKkTENmuFIFTYmethRW};

    %%%%%%%%%%%%
    \nextgroupplot[
        ylabel={EC},
        ylabel style={at={(current axis.east)}, anchor=center, yshift=-1.25em},
    ]
        \addplot[
            matrix plot,  % not matrix plot* 
            draw=black, line width=0.01mm, draw opacity=0.10,
            fill opacity=1,
            colorbar source,
            mesh/cols=19,
            point meta=explicit,
        ]
        table[x=tau_ix_x, y=tau_ix_y, meta=mean_EC] {\SBMOneKkTENmuSIXTYmethRW};
\end{groupplot}

\begin{scope}[shift={(0.95, -2.35)}, scale=1]
    \pgfplotscolorbardrawstandalone[
        colorbar horizontal,
        point meta min=0, point meta max=1,
        colormap name=copper,
        colorbar style={
            height=4pt,
            scale=1,
            xtick style={
                draw=none,
                /pgf/number format/.cd,
                fixed,
                fixed zerofill,
                precision=2,
                /tikz/.cd,
        	},
            xticklabel={\pgfmathparse{\tick}\pgfmathprintnumber[assume math mode=true]{\pgfmathresult}},
            xticklabel style={font=\scriptsize\sffamily, yshift=0.2em},
            xticklabel pos=lower,
            scaled ticks=false,
        },
    ]
\end{scope}
\node[font=\bfseries\large] at (-2em, 7em) {A};
\end{tikzpicture}
%%%%%%%%%%%%%% END of HEATMAPS
% \\[0.5em]
%%%%%%%%%%%% LINE PLOTS 
\begin{tikzpicture}
\begin{groupplot}[
        group style={
		group name=my plots,
		rows=1,
        columns=2,
        % x descriptions at=edge bottom,
        % y descriptions at=edge left,
        xticklabels at=all,
        xlabels at=edge bottom,
        yticklabels at=edge left,
        ylabels at=edge left,
		  vertical sep=20pt,
        horizontal sep=10pt,
	},
        height=120pt,
        width=150pt,
        ymajorgrids,
    	grid style={draw=gray!50, draw opacity=0.75,densely dotted},
        xtick distance=2,
        minor x tick num=1,
        xtick align=center , 
        xtick style={font=\tiny\sffamily},
	ytick style={draw=none, font=\tiny\sffamily},
        xticklabel style={
            font=\tiny\sffamily,
        },
        xtick pos=lower,
        ytick pos=left,
        title style={
            yshift=-1em, 
            align=center, 
            font=\footnotesize\sffamily, 
        },
        xmin=0, xmax=18.5,
        ymin=0, ymax=1,
        % axis background/.style={fill=gray!2},
        axis x line=bottom,
        axis y line=left,
        enlarge x limits=0.02,
        enlarge y limits={value=0.1, upper},
        xlabel={$\tau$-index},
        ylabel style={
            font=\footnotesize\sffamily,
            yshift=-0.25em,
        },
        xlabel style={
            font=\footnotesize\sffamily
        },
        every node near coord/.append style={font=\tiny},
        nodes near coords align={below},
        scaled y ticks=false,
        yticklabel={\pgfmathparse{\tick}\pgfmathprintnumber[assume math mode=true]{\pgfmathresult}},
        legend columns=-1,
        legend pos=outer north east,
        legend style={
            font=\scriptsize,
            column sep=5pt,
            shift={(-8.5em, -20.5em)},
        },
        xtick={},
        scaled ticks=false,
        yticklabel style={
            xshift=0.25em,
            font=\tiny\sffamily,
            /pgf/number format/.cd,
            fixed,
            % fixed zerofill,
            precision=2,
            /tikz/.cd,
    	},
    ]

    \nextgroupplot[ytick distance=0.25, ylabel={EC}]   %%% LFR 1k EC
        \addplot[mu5] table[x=tau_ix, y=EC_mean] {\LFROneKkTENmuFIVEmethRW};
        
        \addplot [name path=upper,draw=none,forget plot] table[x=tau_ix, y=EC_err_max] {\LFROneKkTENmuFIVEmethRW};
        \addplot [name path=lower,draw=none,forget plot] table[x=tau_ix, y=EC_err_min] {\LFROneKkTENmuFIVEmethRW};
        \addplot [mu5_err,forget plot] fill between[of=upper and lower];
        %%%%

        % \addplot[mu10] table[x=tau_ix, y=EC_mean] {\LFROneKkTENmuTENmethRW};
    
        % \addplot [name path=upper,draw=none,forget plot] table[x=tau_ix, y=EC_err_max] {\LFROneKkTENmuTENmethRW};
        % \addplot [name path=lower,draw=none,forget plot] table[x=tau_ix, y=EC_err_min] {\LFROneKkTENmuTENmethRW};
        % \addplot [mu10,forget plot] fill between[of=upper and lower];
        %%%%
        
        \addplot[mu20] table[x=tau_ix, y=EC_mean] {\LFROneKkTENmuTWENTYmethRW};
    
        \addplot [name path=upper,draw=none,forget plot] table[x=tau_ix, y=EC_err_max] {\LFROneKkTENmuTWENTYmethRW};
        \addplot [name path=lower,draw=none,forget plot] table[x=tau_ix, y=EC_err_min] {\LFROneKkTENmuTWENTYmethRW};
        \addplot [mu20_err,forget plot] fill between[of=upper and lower];
        %%%%
        
        % \addplot[mu40] table[x=tau_ix, y=EC_mean] {\LFROneKkTENmuFORTYmethRW};
    
        % \addplot [name path=upper,draw=none,forget plot] table[x=tau_ix, y=EC_err_max] {\LFROneKkTENmuFORTYmethRW};
        % \addplot [name path=lower,draw=none,forget plot] table[x=tau_ix, y=EC_err_min] {\LFROneKkTENmuFORTYmethRW};
        % \addplot [mu40_err,forget plot] fill between[of=upper and lower];
        %%%%
        
        \addplot[mu50] table[x=tau_ix, y=EC_mean] {\LFROneKkTENmuFIFTYmethRW};
        
        \addplot [name path=upper,draw=none,forget plot] table[x=tau_ix, y=EC_err_max] {\LFROneKkTENmuFIFTYmethRW};
        \addplot [name path=lower,draw=none,forget plot] table[x=tau_ix, y=EC_err_min] {\LFROneKkTENmuFIFTYmethRW};
        \addplot [mu50_err,forget plot] fill between[of=upper and lower];
        %%%%
        
        \addplot[mu60] table[x=tau_ix, y=EC_mean] {\LFROneKkTENmuSIXTYmethRW};
    
        \addplot [name path=upper,draw=none,forget plot] table[x=tau_ix, y=EC_err_max] {\LFROneKkTENmuSIXTYmethRW};
        \addplot [name path=lower,draw=none,forget plot] table[x=tau_ix, y=EC_err_min] {\LFROneKkTENmuSIXTYmethRW};
        \addplot [mu60_err,forget plot] fill between[of=upper and lower];
        %%%%

    %%%%%%%%%%%%%%
    \nextgroupplot[ytick distance=0.25,]   %%% SBM 1k RW EC
        \addplot[mu5] table[x=tau_ix, y=EC_mean] {\SBMOneKkTENmuFIVEmethRW};
        
        \addplot [name path=upper,draw=none,forget plot] table[x=tau_ix, y=EC_err_max] {\SBMOneKkTENmuFIVEmethRW};
        \addplot [name path=lower,draw=none,forget plot] table[x=tau_ix, y=EC_err_min] {\SBMOneKkTENmuFIVEmethRW};
        \addplot [mu5_err,forget plot] fill between[of=upper and lower];
        %%%%

        % \addplot[mu10] table[x=tau_ix, y=EC_mean] {\SBMOneKkTENmuTENmethRW};
    
        % \addplot [name path=upper,draw=none,forget plot] table[x=tau_ix, y=EC_err_max] {\SBMOneKkTENmuTENmethRW};
        % \addplot [name path=lower,draw=none,forget plot] table[x=tau_ix, y=EC_err_min] {\SBMOneKkTENmuTENmethRW};
        % \addplot [mu10,forget plot] fill between[of=upper and lower];
        %%%%
        
        \addplot[mu20] table[x=tau_ix, y=EC_mean] {\SBMOneKkTENmuTWENTYmethRW};
    
        \addplot [name path=upper,draw=none,forget plot] table[x=tau_ix, y=EC_err_max] {\SBMOneKkTENmuTWENTYmethRW};
        \addplot [name path=lower,draw=none,forget plot] table[x=tau_ix, y=EC_err_min] {\SBMOneKkTENmuTWENTYmethRW};
        \addplot [mu20_err,forget plot] fill between[of=upper and lower];
        %%%%
        
        % \addplot[mu40] table[x=tau_ix, y=EC_mean] {\SBMOneKkTENmuFORTYmethRW};
    
        % \addplot [name path=upper,draw=none,forget plot] table[x=tau_ix, y=EC_err_max] {\SBMOneKkTENmuFORTYmethRW};
        % \addplot [name path=lower,draw=none,forget plot] table[x=tau_ix, y=EC_err_min] {\SBMOneKkTENmuFORTYmethRW};
        % \addplot [mu40_err,forget plot] fill between[of=upper and lower];
        %%%%
        
        \addplot[mu50] table[x=tau_ix, y=EC_mean] {\SBMOneKkTENmuFIFTYmethRW};
        
        \addplot [name path=upper,draw=none,forget plot] table[x=tau_ix, y=EC_err_max] {\SBMOneKkTENmuFIFTYmethRW};
        \addplot [name path=lower,draw=none,forget plot] table[x=tau_ix, y=EC_err_min] {\SBMOneKkTENmuFIFTYmethRW};
        \addplot [mu50_err,forget plot] fill between[of=upper and lower];
        %%%%
        
        \addplot[mu60] table[x=tau_ix, y=EC_mean] {\SBMOneKkTENmuSIXTYmethRW};
    
        \addplot [name path=upper,draw=none,forget plot] table[x=tau_ix, y=EC_err_max] {\SBMOneKkTENmuSIXTYmethRW};
        \addplot [name path=lower,draw=none,forget plot] table[x=tau_ix, y=EC_err_min] {\SBMOneKkTENmuSIXTYmethRW};
        \addplot [mu60_err,forget plot] fill between[of=upper and lower];
        %%%%

\end{groupplot}

\node[font=\bfseries\large] at (-2em, 9em) {B};

\end{tikzpicture}

%% file: new-figs/inertia-heatmaps.tex
\input{data/inertia-heatmap-data}
% \pgfplotsset{colormap/GnBu}
\pgfplotsset{colormap/viridis high res}

\pgfplotsset{
  coordinate style/.condition={0}{},
}

\begin{tikzpicture}[scale=0.95, transform shape]
    \begin{groupplot}[
        group style={
            rows=2,
            columns=3,
            group name=mygroup,
            vertical sep=15pt,
            horizontal sep=25pt,
            % x descriptions at=edge bottom,
            % xlabels at=edge top,
            xticklabels at=edge top,
            y descriptions at=edge left,
        },
        % width=0.50\textwidth,
        height=140pt, width=135pt,
        scale only axis=true,
        axis on top,
        % axis equal image,
        % colormap name=copper,
        colormap name=viridis high res,
        % y dir=reverse,
        ytick=data,
        xtick=data,
        symbolic x coords={0.05,0.1,0.2,0.4,0.5,0.6},
        symbolic y coords={0,1,2,3,4,5,10,20,50,100,200,500,750,1000,2000,5000},
        enlarge x limits=false,
        enlarge y limits=false,
        xmin={[normalized]-0.50}, xmax={[normalized]5.50},
        ymin={[normalized]-0.50}, ymax={[normalized]15.50},
        % xmin=0.05, xmax=5.5,
        % ymin=-0.5, ymax=5000,
        % xtick distance=1, ytick distance=1,
        % xtick distance=2, ytick distance=2,
        % xtick distance=3, ytick distance=3,
        typeset ticklabels with strut,
        nodes near coords={%
            \pgfmathfloatparsenumber{\pgfplotspointmeta}
            \pgfmathfloatgetflagstomacro\pgfmathresult\currentflags
            \ifnum\currentflags=3\else
                \pgfmathprintnumber\pgfplotspointmeta
            \fi
        },   % this filters out nans
        nodes near coords align={anchor=center},
        nodes near coords style={
            font=\fontsize{2pt}{2pt}\sffamily\selectfont,
            /pgf/number format/.cd,
            fixed, %fixed zerofill, 
            precision=2,
            /tikz/.cd,
        },
        % nodes near coords style={font=\tiny},
        % yticklabel={\pgfmathparse{\tick*100}\pgfmathprintnumber[assume math mode=true]{\pgfmathresult}\%},
        title style={align=center, yshift=0.5em, font=\small\sffamily, },
        ylabel style={font=\small\sffamily, yshift=-0em},
        xlabel style={font=\small\sffamily, yshift=-0em},
        % colorbar
        tick align=outside,
        xticklabel pos=top,
        xtick style={draw=none},
        ytick style={draw=none},
        xticklabel style={draw=none, font=\scriptsize\sffamily, yshift=-0.7em},
        yticklabel style={draw=none, font=\scriptsize\sffamily, xshift=0.35em},
        % axis background/.style={fill=gray!90},
        axis line style={draw=black},
        point meta min=0,
        point meta max=1,
        clip=false,
        unbounded coords=discard,
        axis background/.style={fill=gray!30},
    ]
    %%%%%%%%% LFR-1k, avg_k = 10
    \nextgroupplot[
        title={$\left< k \right> = 10$},
        xlabel={$\mu$},
        xlabel style={xshift=-8.5em, yshift=-1.3em, font=\small\sffamily},
        ylabel={Inertia},
    ]
        \addplot[
            matrix plot,  % not matrix plot* 
            draw=black, line width=0.05mm, draw opacity=0.2, 
            fill opacity=1,
            colorbar source,
            mesh/cols=6,
            point meta=explicit,
            coordinate style/.condition={meta<0.60}{white},
        ]
        table[
            x=mu, y=inertia, meta=EC,
        ] 
        {\LFROneKkTENmethRWEC};

    %%%%%%%%% LFR-1k EC, avg_k = 15
    \nextgroupplot[
        title={$\left< k \right> = 15$},
    ]
        \addplot[
            matrix plot,  % not matrix plot* 
            draw=black, line width=0.05mm, draw opacity=0.2, 
            fill opacity=1,
            colorbar source,
            mesh/cols=6,
            point meta=explicit,
            coordinate style/.condition={meta<0.60}{white},
        ]
        table[
            x=mu, y=inertia, meta=EC,
        ] 
        {\LFROneKkFIFTEENmethRWEC};

    %%%%%%%%% LFR-1k EC, avg_k = 20
    \nextgroupplot[
        title={$\left< k \right> = 20$},
        ylabel={EC},
        ylabel style={at={(current axis.east)}, anchor=center, yshift=-2.75em},
    ]
        \addplot[
            matrix plot,  % not matrix plot* 
            draw=black, line width=0.05mm, draw opacity=0.2, 
            fill opacity=1,
            colorbar source,
            mesh/cols=6,
            point meta=explicit,
            coordinate style/.condition={meta<0.60}{white},
        ]
        table[
            x=mu, y=inertia, meta=EC,
        ] 
        {\LFROneKkTWENTYmethRWEC};
    %%%%%%%%%%%%%%%%%%%%%%%

    %% this row for EC improvement
    \nextgroupplot[
        ylabel={Inertia},
        ymin={[normalized]0.5},
        clip=true,
    ]
        \addplot[
            matrix plot,  % not matrix plot* 
            draw=black, line width=0.05mm, draw opacity=0.2, 
            fill opacity=1,
            colorbar source,
            mesh/cols=6,
            point meta=explicit,
            coordinate style/.condition={meta<0.60}{white},
        ]
        table[
            x=mu, y=inertia, meta=EC_improv,
        ] 
        {\LFROneKkTENmethRWEC};

    \nextgroupplot[
        ymin={[normalized]0.5},
        clip=true,
    ]
        \addplot[
            matrix plot,  % not matrix plot* 
            draw=black, line width=0.05mm, draw opacity=0.2, 
            fill opacity=1,
            colorbar source,
            mesh/cols=6,
            point meta=explicit,
            coordinate style/.condition={meta<0.60}{white},
        ]
        table[
            x=mu, y=inertia, meta=EC_improv,
        ] 
        {\LFROneKkFIFTEENmethRWEC};

    \nextgroupplot[
        ylabel={Relative Improvement \\ over RW w/o inertia},
        ylabel style={at={(current axis.east)}, anchor=center, yshift=-2.75em, align=center},
        ymin={[normalized]0.5},
        clip=true,
    ]
        \addplot[
            matrix plot,  % not matrix plot* 
            draw=black, line width=0.05mm, draw opacity=0.2, 
            fill opacity=1,
            colorbar source,
            mesh/cols=6,
            point meta=explicit,
            coordinate style/.condition={meta<0.60}{white},
        ]
        table[
            x=mu, y=inertia, meta=EC_improv,
        ] 
        {\LFROneKkFIFTEENmethRWEC};
    
    \end{groupplot}

\begin{scope}[shift={(4, -5.65)}, scale=1]
    \pgfplotscolorbardrawstandalone[
        colorbar horizontal,
        point meta min=0, point meta max=1,
        colormap name=viridis high res,
        colorbar style={
            height=5pt,
            scale=1,
            xtick style={
                draw=none,
                /pgf/number format/.cd,
                fixed,
                fixed zerofill,
                precision=0,
                /tikz/.cd,
        	},
            xticklabel={\pgfmathparse{\tick*100}\pgfmathprintnumber[assume math mode=true]{\pgfmathresult}\%},
            xticklabel={\pgfmathparse{\tick*1}\pgfmathprintnumber[assume math mode=true]{\pgfmathresult}},
            xticklabel style={font=\scriptsize\sffamily, yshift=0.2em},
            xticklabel pos=lower,
            scaled ticks=false,
        },
    ]
\end{scope}
\end{tikzpicture}

%% file: new-figs/methods-comparison-heatmap.tex
\input{data/method-comparison-heatmap-data}
% \pgfplotsset{colormap/GnBu}
% \pgfplotsset{colormap/copper}
\pgfplotsset{colormap/viridis high res}

\pgfplotsset{
  coordinate style/.condition={0}{},
}

\begin{tikzpicture}[scale=0.95, transform shape]
    \begin{groupplot}[
        group style={
            rows=2,
            columns=3,
            group name=mygroup,
            vertical sep=15pt,
            horizontal sep=25pt,
            % x descriptions at=edge bottom,
            % xlabels at=edge top,
            xticklabels at=edge top,
            y descriptions at=edge left,
        },
        % width=0.50\textwidth,
        height=165pt, width=135pt,
        scale only axis=true,
        axis on top,
        % axis equal image,
        colormap name=viridis high res,
        % y dir=reverse,
        ytick=data,
        xtick=data,
        symbolic x coords={0.05,0.1,0.2,0.4,0.5,0.6},
        symbolic y coords={Markov-Diff,Markov-RW,MultiLouvain,RW,RW-1,RW-2,RW-3,RW-4,RW-5,RW-10,RW-20,RW-50,RW-100,RW-200,RW-500,RW-750,RW-1000,RW-2000,RW-5000},
        enlarge x limits=false,
        enlarge y limits=false,
        xmin={[normalized]-0.50}, xmax={[normalized]5.50},
        ymin={[normalized]-0.50}, ymax={[normalized]18.50},
        % xmin=0.05, xmax=5.5,
        % ymin=-0.5, ymax=5000,
        % xtick distance=1, ytick distance=1,
        % xtick distance=2, ytick distance=2,
        % xtick distance=3, ytick distance=3,
        typeset ticklabels with strut,
        nodes near coords={%
            \pgfmathfloatparsenumber{\pgfplotspointmeta}
            \pgfmathfloatgetflagstomacro\pgfmathresult\currentflags
            \ifnum\currentflags=3\else
                \pgfmathprintnumber\pgfplotspointmeta
            \fi
        },   % this filters out nans
        nodes near coords align={anchor=center},
        nodes near coords style={
            font=\tiny,
            /pgf/number format/.cd,
            fixed, %fixed zerofill, 
            precision=2,
            /tikz/.cd,
        },
        % nodes near coords style={font=\tiny},
        % yticklabel={\pgfmathparse{\tick*100}\pgfmathprintnumber[assume math mode=true]{\pgfmathresult}\%},
        title style={align=center, yshift=1em, font=\small\sffamily, },
        ylabel style={font=\small\sffamily, yshift=-0em},
        xlabel style={font=\small\sffamily, yshift=-0em},
        % colorbar
        tick align=outside,
        xticklabel pos=top,
        xtick style={draw=none},
        ytick style={draw=none},
        xticklabel style={draw=none, font=\scriptsize\sffamily, yshift=-0.5em},
        yticklabel style={draw=none, font=\scriptsize\sffamily, xshift=0.35em},
        % axis background/.style={fill=gray!90},
        axis line style={draw=black},
        point meta min=0,
        point meta max=1,
        clip=false,
        unbounded coords=discard,
        axis background/.style={fill=gray!30},
    ]
    %%%%%%%%% LFR-1k, avg_k = 10
    \nextgroupplot[
        title={$\left< k \right> = 10$},
        xlabel={$\mu$},
        xlabel style={xshift=-8.5em, yshift=-1.3em, font=\small\sffamily},
        ylabel={LFR},
    ]
        \addplot[
            matrix plot,  % not matrix plot* 
            draw=black, line width=0.05mm, draw opacity=0.2, 
            fill opacity=1,
            colorbar source,
            mesh/cols=6,
            point meta=explicit,
            coordinate style/.condition={meta<0.60}{white},
        ]
        table[
            x=mu, y=method, meta=EC,
        ] 
        {\LFROneKkTENallmethodsEC};

    % %%%%%%%%% LFR-1k EC, avg_k = 15
    \nextgroupplot[
        title={$\left< k \right> = 15$},
    ]
        \addplot[
            matrix plot,  % not matrix plot* 
            draw=black, line width=0.05mm, draw opacity=0.2, 
            fill opacity=1,
            colorbar source,
            mesh/cols=6,
            point meta=explicit,
            coordinate style/.condition={meta<0.60}{white},
        ]
        table[
            x=mu, y=method, meta=EC,
        ] 
        {\LFROneKkFIFTEENallmethodsEC};

    %%%%%%%%% LFR-1k EC, avg_k = 20
    \nextgroupplot[
        title={$\left< k \right> = 20$},
        ylabel={EC},
        ylabel style={at={(current axis.east)}, anchor=center, yshift=-2.75em},
    ]
        \addplot[
            matrix plot,  % not matrix plot* 
            draw=black, line width=0.05mm, draw opacity=0.2, 
            fill opacity=1,
            colorbar source,
            mesh/cols=6,
            point meta=explicit,
            coordinate style/.condition={meta<0.60}{white},
        ]
        table[
            x=mu, y=method, meta=EC,
        ] 
        {\LFROneKkTWENTYallmethodsEC};
    % %%%%%%%%%%%%%%%%%%%%%%%
    %%%%%%%%% SBM-1k, avg_k = 10
    \nextgroupplot[
        ylabel={SBM},
    ]
        \addplot[
            matrix plot,  % not matrix plot* 
            draw=black, line width=0.05mm, draw opacity=0.2, 
            fill opacity=1,
            colorbar source,
            mesh/cols=6,
            point meta=explicit,
            coordinate style/.condition={meta<0.60}{white},
        ]
        table[
            x=mu, y=method, meta=EC,
        ] 
        {\SBMOneKkTENallmethodsEC};

    % %%%%%%%%% SBM-1k EC, avg_k = 15
    \nextgroupplot[
    ]
        \addplot[
            matrix plot,  % not matrix plot* 
            draw=black, line width=0.05mm, draw opacity=0.2, 
            fill opacity=1,
            colorbar source,
            mesh/cols=6,
            point meta=explicit,
            coordinate style/.condition={meta<0.60}{white},
        ]
        table[
            x=mu, y=method, meta=EC,
        ] 
        {\SBMOneKkFIFTEENallmethodsEC};

    %%%%%%%%% SBM-1k EC, avg_k = 20
    \nextgroupplot[
        ylabel={EC},
        ylabel style={at={(current axis.east)}, anchor=center, yshift=-2.5em},
    ]
        \addplot[
            matrix plot,  % not matrix plot* 
            draw=black, line width=0.05mm, draw opacity=0.2, 
            fill opacity=1,
            colorbar source,
            mesh/cols=6,
            point meta=explicit,
            coordinate style/.condition={meta<0.60}{white},
        ]
        table[
            x=mu, y=method, meta=EC,
        ] 
        {\SBMOneKkTWENTYallmethodsEC};
    % %%%%%%%%%%%%%%%%%%%%%%%
    \end{groupplot}

\begin{scope}[shift={(4, -6.7)}, scale=1]
    \pgfplotscolorbardrawstandalone[
        colorbar horizontal,
        point meta min=0, point meta max=1,
        colormap name=viridis high res,
        colorbar style={
            height=5pt,
            scale=1,
            xtick style={
                draw=none,
                /pgf/number format/.cd,
                fixed,
                fixed zerofill,
                precision=0,
                /tikz/.cd,
        	},
            xticklabel={\pgfmathparse{\tick*1}\pgfmathprintnumber[assume math mode=true]{\pgfmathresult}},
            xticklabel style={font=\scriptsize\sffamily, yshift=0.2em},
            xticklabel pos=lower,
            scaled ticks=false,
        },
    ]
\end{scope}
\end{tikzpicture}

%% file: data/method-comparison-heatmap-data.tex
\pgfplotstableread[ignore chars={"}]{
    method mu EC avg_k family inertia
    Markov-Diff 0.05 1 10 LFROneK 0
    Markov-Diff 0.1 1 10 LFROneK 0
    Markov-Diff 0.2 1 10 LFROneK 0
    Markov-Diff 0.4 0.957 10 LFROneK 0
    Markov-Diff 0.5 0.8572 10 LFROneK 0
    Markov-Diff 0.6 0.5579 10 LFROneK 0
    Markov-RW 0.05 1 10 LFROneK 0
    Markov-RW 0.1 1 10 LFROneK 0
    Markov-RW 0.2 1 10 LFROneK 0
    Markov-RW 0.4 0.9859 10 LFROneK 0
    Markov-RW 0.5 0.9401 10 LFROneK 0
    Markov-RW 0.6 0.6535 10 LFROneK 0
    MultiLouvain 0.05 1 10 LFROneK 0
    MultiLouvain 0.1 1 10 LFROneK 0
    MultiLouvain 0.2 1 10 LFROneK 0
    MultiLouvain 0.4 0.9888 10 LFROneK 0
    MultiLouvain 0.5 0.9251 10 LFROneK 0
    MultiLouvain 0.6 0.6776 10 LFROneK 0
    RW 0.05 1 10 LFROneK 0
    RW 0.1 0.84 10 LFROneK 0
    RW 0.2 0.747 10 LFROneK 0
    RW 0.4 0.591 10 LFROneK 0
    RW 0.5 0.4334 10 LFROneK 0
    RW 0.6 0.1719 10 LFROneK 0
    RW-1 0.05 1 10 LFROneK 1
    RW-1 0.1 0.9164 10 LFROneK 1
    RW-1 0.2 0.8972 10 LFROneK 1
    RW-1 0.4 0.7966 10 LFROneK 1
    RW-1 0.5 0.6509 10 LFROneK 1
    RW-1 0.6 0.204 10 LFROneK 1
    RW-2 0.05 1 10 LFROneK 2
    RW-2 0.1 0.9861 10 LFROneK 2
    RW-2 0.2 0.9771 10 LFROneK 2
    RW-2 0.4 0.8722 10 LFROneK 2
    RW-2 0.5 0.6471 10 LFROneK 2
    RW-2 0.6 0.2253 10 LFROneK 2
    RW-3 0.05 1 10 LFROneK 3
    RW-3 0.1 1 10 LFROneK 3
    RW-3 0.2 1 10 LFROneK 3
    RW-3 0.4 0.8738 10 LFROneK 3
    RW-3 0.5 0.685 10 LFROneK 3
    RW-3 0.6 0.2334 10 LFROneK 3
    RW-4 0.05 1 10 LFROneK 4
    RW-4 0.1 1 10 LFROneK 4
    RW-4 0.2 0.9971 10 LFROneK 4
    RW-4 0.4 0.8962 10 LFROneK 4
    RW-4 0.5 0.6565 10 LFROneK 4
    RW-4 0.6 0.2388 10 LFROneK 4
    RW-5 0.05 1 10 LFROneK 5
    RW-5 0.1 1 10 LFROneK 5
    RW-5 0.2 0.9891 10 LFROneK 5
    RW-5 0.4 0.8779 10 LFROneK 5
    RW-5 0.5 0.6506 10 LFROneK 5
    RW-5 0.6 0.2374 10 LFROneK 5
    RW-10 0.05 1 10 LFROneK 10
    RW-10 0.1 0.9861 10 LFROneK 10
    RW-10 0.2 0.9971 10 LFROneK 10
    RW-10 0.4 0.8923 10 LFROneK 10
    RW-10 0.5 0.6519 10 LFROneK 10
    RW-10 0.6 0.2323 10 LFROneK 10
    RW-20 0.05 1 10 LFROneK 20
    RW-20 0.1 1 10 LFROneK 20
    RW-20 0.2 0.9891 10 LFROneK 20
    RW-20 0.4 0.8727 10 LFROneK 20
    RW-20 0.5 0.6574 10 LFROneK 20
    RW-20 0.6 0.2302 10 LFROneK 20
    RW-50 0.05 1 10 LFROneK 50
    RW-50 0.1 1 10 LFROneK 50
    RW-50 0.2 1 10 LFROneK 50
    RW-50 0.4 0.8873 10 LFROneK 50
    RW-50 0.5 0.6673 10 LFROneK 50
    RW-50 0.6 0.2394 10 LFROneK 50
    RW-100 0.05 1 10 LFROneK 100
    RW-100 0.1 1 10 LFROneK 100
    RW-100 0.2 1 10 LFROneK 100
    RW-100 0.4 0.8857 10 LFROneK 100
    RW-100 0.5 0.653 10 LFROneK 100
    RW-100 0.6 0.2387 10 LFROneK 100
    RW-200 0.05 1 10 LFROneK 200
    RW-200 0.1 1 10 LFROneK 200
    RW-200 0.2 1 10 LFROneK 200
    RW-200 0.4 0.8841 10 LFROneK 200
    RW-200 0.5 0.6571 10 LFROneK 200
    RW-200 0.6 0.2363 10 LFROneK 200
    RW-500 0.05 1 10 LFROneK 500
    RW-500 0.1 1 10 LFROneK 500
    RW-500 0.2 1 10 LFROneK 500
    RW-500 0.4 0.8804 10 LFROneK 500
    RW-500 0.5 0.6431 10 LFROneK 500
    RW-500 0.6 0.2345 10 LFROneK 500
    RW-750 0.05 1 10 LFROneK 750
    RW-750 0.1 1 10 LFROneK 750
    RW-750 0.2 1 10 LFROneK 750
    RW-750 0.4 0.8871 10 LFROneK 750
    RW-750 0.5 0.6757 10 LFROneK 750
    RW-750 0.6 0.2328 10 LFROneK 750
    RW-1000 0.05 1 10 LFROneK 1000
    RW-1000 0.1 1 10 LFROneK 1000
    RW-1000 0.2 1 10 LFROneK 1000
    RW-1000 0.4 0.8858 10 LFROneK 1000
    RW-1000 0.5 0.6475 10 LFROneK 1000
    RW-1000 0.6 0.2292 10 LFROneK 1000
    RW-2000 0.05 1 10 LFROneK 2000
    RW-2000 0.1 1 10 LFROneK 2000
    RW-2000 0.2 1 10 LFROneK 2000
    RW-2000 0.4 0.8873 10 LFROneK 2000
    RW-2000 0.5 0.6442 10 LFROneK 2000
    RW-2000 0.6 0.235 10 LFROneK 2000
    RW-5000 0.05 1 10 LFROneK 5000
    RW-5000 0.1 1 10 LFROneK 5000
    RW-5000 0.2 1 10 LFROneK 5000
    RW-5000 0.4 0.8855 10 LFROneK 5000
    RW-5000 0.5 0.6275 10 LFROneK 5000
    RW-5000 0.6 0.2268 10 LFROneK 5000
}{\LFROneKkTENallmethodsEC}

%%%%%%%%%%%%%%%%%

\pgfplotstableread[ignore chars={"}]{
    method mu EC avg_k family inertia
    Markov-Diff 0.05 1 15 LFROneK 0
    Markov-Diff 0.1 1 15 LFROneK 0
    Markov-Diff 0.2 1 15 LFROneK 0
    Markov-Diff 0.4 1 15 LFROneK 0
    Markov-Diff 0.5 0.9971 15 LFROneK 0
    Markov-Diff 0.6 0.9483 15 LFROneK 0
    Markov-RW 0.05 1 15 LFROneK 0
    Markov-RW 0.1 1 15 LFROneK 0
    Markov-RW 0.2 1 15 LFROneK 0
    Markov-RW 0.4 1 15 LFROneK 0
    Markov-RW 0.5 0.9943 15 LFROneK 0
    Markov-RW 0.6 0.978 15 LFROneK 0
    MultiLouvain 0.05 1 15 LFROneK 0
    MultiLouvain 0.1 1 15 LFROneK 0
    MultiLouvain 0.2 1 15 LFROneK 0
    MultiLouvain 0.4 1 15 LFROneK 0
    MultiLouvain 0.5 0.9943 15 LFROneK 0
    MultiLouvain 0.6 0.9764 15 LFROneK 0
    RW 0.05 1 15 LFROneK 0
    RW 0.1 0.8533 15 LFROneK 0
    RW 0.2 0.7972 15 LFROneK 0
    RW 0.4 0.6867 15 LFROneK 0
    RW 0.5 0.5865 15 LFROneK 0
    RW 0.6 0.4703 15 LFROneK 0
    RW-1 0.05 1 15 LFROneK 1
    RW-1 0.1 0.9795 15 LFROneK 1
    RW-1 0.2 0.9552 15 LFROneK 1
    RW-1 0.4 0.9138 15 LFROneK 1
    RW-1 0.5 0.9147 15 LFROneK 1
    RW-1 0.6 0.8232 15 LFROneK 1
    RW-2 0.05 1 15 LFROneK 2
    RW-2 0.1 1 15 LFROneK 2
    RW-2 0.2 0.989 15 LFROneK 2
    RW-2 0.4 1 15 LFROneK 2
    RW-2 0.5 0.9625 15 LFROneK 2
    RW-2 0.6 0.872 15 LFROneK 2
    RW-3 0.05 1 15 LFROneK 3
    RW-3 0.1 1 15 LFROneK 3
    RW-3 0.2 1 15 LFROneK 3
    RW-3 0.4 0.9971 15 LFROneK 3
    RW-3 0.5 0.9914 15 LFROneK 3
    RW-3 0.6 0.8573 15 LFROneK 3
    RW-4 0.05 1 15 LFROneK 4
    RW-4 0.1 1 15 LFROneK 4
    RW-4 0.2 1 15 LFROneK 4
    RW-4 0.4 1 15 LFROneK 4
    RW-4 0.5 0.9818 15 LFROneK 4
    RW-4 0.6 0.8588 15 LFROneK 4
    RW-5 0.05 1 15 LFROneK 5
    RW-5 0.1 1 15 LFROneK 5
    RW-5 0.2 1 15 LFROneK 5
    RW-5 0.4 1 15 LFROneK 5
    RW-5 0.5 0.9943 15 LFROneK 5
    RW-5 0.6 0.8859 15 LFROneK 5
    RW-10 0.05 1 15 LFROneK 10
    RW-10 0.1 1 15 LFROneK 10
    RW-10 0.2 0.9971 15 LFROneK 10
    RW-10 0.4 1 15 LFROneK 10
    RW-10 0.5 0.9723 15 LFROneK 10
    RW-10 0.6 0.8671 15 LFROneK 10
    RW-20 0.05 1 15 LFROneK 20
    RW-20 0.1 1 15 LFROneK 20
    RW-20 0.2 1 15 LFROneK 20
    RW-20 0.4 1 15 LFROneK 20
    RW-20 0.5 0.9859 15 LFROneK 20
    RW-20 0.6 0.8655 15 LFROneK 20
    RW-50 0.05 1 15 LFROneK 50
    RW-50 0.1 1 15 LFROneK 50
    RW-50 0.2 1 15 LFROneK 50
    RW-50 0.4 1 15 LFROneK 50
    RW-50 0.5 0.9943 15 LFROneK 50
    RW-50 0.6 0.8374 15 LFROneK 50
    RW-100 0.05 1 15 LFROneK 100
    RW-100 0.1 1 15 LFROneK 100
    RW-100 0.2 1 15 LFROneK 100
    RW-100 0.4 1 15 LFROneK 100
    RW-100 0.5 0.9914 15 LFROneK 100
    RW-100 0.6 0.8434 15 LFROneK 100
    RW-200 0.05 1 15 LFROneK 200
    RW-200 0.1 1 15 LFROneK 200
    RW-200 0.2 1 15 LFROneK 200
    RW-200 0.4 1 15 LFROneK 200
    RW-200 0.5 0.9914 15 LFROneK 200
    RW-200 0.6 0.8425 15 LFROneK 200
    RW-500 0.05 1 15 LFROneK 500
    RW-500 0.1 1 15 LFROneK 500
    RW-500 0.2 1 15 LFROneK 500
    RW-500 0.4 0.9971 15 LFROneK 500
    RW-500 0.5 0.9914 15 LFROneK 500
    RW-500 0.6 0.838 15 LFROneK 500
    RW-750 0.05 1 15 LFROneK 750
    RW-750 0.1 1 15 LFROneK 750
    RW-750 0.2 1 15 LFROneK 750
    RW-750 0.4 0.9971 15 LFROneK 750
    RW-750 0.5 0.9914 15 LFROneK 750
    RW-750 0.6 0.8416 15 LFROneK 750
    RW-1000 0.05 1 15 LFROneK 1000
    RW-1000 0.1 1 15 LFROneK 1000
    RW-1000 0.2 1 15 LFROneK 1000
    RW-1000 0.4 0.9971 15 LFROneK 1000
    RW-1000 0.5 0.9914 15 LFROneK 1000
    RW-1000 0.6 0.8401 15 LFROneK 1000
    RW-2000 0.05 1 15 LFROneK 2000
    RW-2000 0.1 1 15 LFROneK 2000
    RW-2000 0.2 1 15 LFROneK 2000
    RW-2000 0.4 0.9971 15 LFROneK 2000
    RW-2000 0.5 0.9886 15 LFROneK 2000
    RW-2000 0.6 0.8483 15 LFROneK 2000
    RW-5000 0.05 1 15 LFROneK 5000
    RW-5000 0.1 1 15 LFROneK 5000
    RW-5000 0.2 1 15 LFROneK 5000
    RW-5000 0.4 0.9971 15 LFROneK 5000
    RW-5000 0.5 0.9914 15 LFROneK 5000
    RW-5000 0.6 0.855 15 LFROneK 5000
}{\LFROneKkFIFTEENallmethodsEC}

%%%%%%%%%%%%%%%%%

\pgfplotstableread[ignore chars={"}]{
    method mu EC avg_k family inertia
    Markov-Diff 0.05 1 20 LFROneK 0
    Markov-Diff 0.1 1 20 LFROneK 0
    Markov-Diff 0.2 1 20 LFROneK 0
    Markov-Diff 0.4 1 20 LFROneK 0
    Markov-Diff 0.5 1 20 LFROneK 0
    Markov-Diff 0.6 1 20 LFROneK 0
    Markov-RW 0.05 1 20 LFROneK 0
    Markov-RW 0.1 1 20 LFROneK 0
    Markov-RW 0.2 1 20 LFROneK 0
    Markov-RW 0.4 1 20 LFROneK 0
    Markov-RW 0.5 1 20 LFROneK 0
    Markov-RW 0.6 1 20 LFROneK 0
    MultiLouvain 0.05 1 20 LFROneK 0
    MultiLouvain 0.1 1 20 LFROneK 0
    MultiLouvain 0.2 1 20 LFROneK 0
    MultiLouvain 0.4 1 20 LFROneK 0
    MultiLouvain 0.5 1 20 LFROneK 0
    MultiLouvain 0.6 1 20 LFROneK 0
    RW 0.05 1 20 LFROneK 0
    RW 0.1 0.854 20 LFROneK 0
    RW 0.2 0.7797 20 LFROneK 0
    RW 0.4 0.6837 20 LFROneK 0
    RW 0.5 0.6741 20 LFROneK 0
    RW 0.6 0.5618 20 LFROneK 0
    RW-1 0.05 1 20 LFROneK 1
    RW-1 0.1 0.9881 20 LFROneK 1
    RW-1 0.2 0.9632 20 LFROneK 1
    RW-1 0.4 0.9304 20 LFROneK 1
    RW-1 0.5 0.9183 20 LFROneK 1
    RW-1 0.6 0.8938 20 LFROneK 1
    RW-2 0.05 1 20 LFROneK 2
    RW-2 0.1 1 20 LFROneK 2
    RW-2 0.2 1 20 LFROneK 2
    RW-2 0.4 1 20 LFROneK 2
    RW-2 0.5 1 20 LFROneK 2
    RW-2 0.6 1 20 LFROneK 2
    RW-3 0.05 1 20 LFROneK 3
    RW-3 0.1 1 20 LFROneK 3
    RW-3 0.2 1 20 LFROneK 3
    RW-3 0.4 1 20 LFROneK 3
    RW-3 0.5 1 20 LFROneK 3
    RW-3 0.6 0.9971 20 LFROneK 3
    RW-4 0.05 1 20 LFROneK 4
    RW-4 0.1 1 20 LFROneK 4
    RW-4 0.2 1 20 LFROneK 4
    RW-4 0.4 1 20 LFROneK 4
    RW-4 0.5 1 20 LFROneK 4
    RW-4 0.6 0.9971 20 LFROneK 4
    RW-5 0.05 1 20 LFROneK 5
    RW-5 0.1 1 20 LFROneK 5
    RW-5 0.2 1 20 LFROneK 5
    RW-5 0.4 1 20 LFROneK 5
    RW-5 0.5 1 20 LFROneK 5
    RW-5 0.6 0.9971 20 LFROneK 5
    RW-10 0.05 1 20 LFROneK 10
    RW-10 0.1 1 20 LFROneK 10
    RW-10 0.2 1 20 LFROneK 10
    RW-10 0.4 1 20 LFROneK 10
    RW-10 0.5 1 20 LFROneK 10
    RW-10 0.6 0.9971 20 LFROneK 10
    RW-20 0.05 1 20 LFROneK 20
    RW-20 0.1 1 20 LFROneK 20
    RW-20 0.2 1 20 LFROneK 20
    RW-20 0.4 1 20 LFROneK 20
    RW-20 0.5 1 20 LFROneK 20
    RW-20 0.6 0.9971 20 LFROneK 20
    RW-50 0.05 1 20 LFROneK 50
    RW-50 0.1 1 20 LFROneK 50
    RW-50 0.2 1 20 LFROneK 50
    RW-50 0.4 1 20 LFROneK 50
    RW-50 0.5 1 20 LFROneK 50
    RW-50 0.6 0.9971 20 LFROneK 50
    RW-100 0.05 1 20 LFROneK 100
    RW-100 0.1 1 20 LFROneK 100
    RW-100 0.2 1 20 LFROneK 100
    RW-100 0.4 1 20 LFROneK 100
    RW-100 0.5 1 20 LFROneK 100
    RW-100 0.6 0.9971 20 LFROneK 100
    RW-200 0.05 1 20 LFROneK 200
    RW-200 0.1 1 20 LFROneK 200
    RW-200 0.2 1 20 LFROneK 200
    RW-200 0.4 1 20 LFROneK 200
    RW-200 0.5 1 20 LFROneK 200
    RW-200 0.6 0.9971 20 LFROneK 200
    RW-500 0.05 1 20 LFROneK 500
    RW-500 0.1 1 20 LFROneK 500
    RW-500 0.2 1 20 LFROneK 500
    RW-500 0.4 1 20 LFROneK 500
    RW-500 0.5 1 20 LFROneK 500
    RW-500 0.6 0.9971 20 LFROneK 500
    RW-750 0.05 1 20 LFROneK 750
    RW-750 0.1 1 20 LFROneK 750
    RW-750 0.2 1 20 LFROneK 750
    RW-750 0.4 1 20 LFROneK 750
    RW-750 0.5 1 20 LFROneK 750
    RW-750 0.6 0.9971 20 LFROneK 750
    RW-1000 0.05 1 20 LFROneK 1000
    RW-1000 0.1 1 20 LFROneK 1000
    RW-1000 0.2 1 20 LFROneK 1000
    RW-1000 0.4 1 20 LFROneK 1000
    RW-1000 0.5 1 20 LFROneK 1000
    RW-1000 0.6 0.9971 20 LFROneK 1000
    RW-2000 0.05 1 20 LFROneK 2000
    RW-2000 0.1 1 20 LFROneK 2000
    RW-2000 0.2 1 20 LFROneK 2000
    RW-2000 0.4 1 20 LFROneK 2000
    RW-2000 0.5 1 20 LFROneK 2000
    RW-2000 0.6 0.9971 20 LFROneK 2000
    RW-5000 0.05 1 20 LFROneK 5000
    RW-5000 0.1 1 20 LFROneK 5000
    RW-5000 0.2 1 20 LFROneK 5000
    RW-5000 0.4 1 20 LFROneK 5000
    RW-5000 0.5 1 20 LFROneK 5000
    RW-5000 0.6 0.9971 20 LFROneK 5000
}{\LFROneKkTWENTYallmethodsEC}

\pgfplotstableread[ignore chars={"}]{
    method mu EC avg_k family inertia
    Markov-Diff 0.05 1 10 SBMOneK 0
    Markov-Diff 0.1 1 10 SBMOneK 0
    Markov-Diff 0.2 1 10 SBMOneK 0
    Markov-Diff 0.4 0.9748 10 SBMOneK 0
    Markov-Diff 0.5 0.8784 10 SBMOneK 0
    Markov-Diff 0.6 0.3995 10 SBMOneK 0
    Markov-RW 0.05 1 10 SBMOneK 0
    Markov-RW 0.1 1 10 SBMOneK 0
    Markov-RW 0.2 1 10 SBMOneK 0
    Markov-RW 0.4 0.9708 10 SBMOneK 0
    Markov-RW 0.5 0.8903 10 SBMOneK 0
    Markov-RW 0.6 0.4859 10 SBMOneK 0
    MultiLouvain 0.05 1 10 SBMOneK 0
    MultiLouvain 0.1 1 10 SBMOneK 0
    MultiLouvain 0.2 1 10 SBMOneK 0
    MultiLouvain 0.4 0.9631 10 SBMOneK 0
    MultiLouvain 0.5 0.7739 10 SBMOneK 0
    MultiLouvain 0.6 0.3001 10 SBMOneK 0
    RW 0.05 1 10 SBMOneK 0
    RW 0.1 1 10 SBMOneK 0
    RW 0.2 1 10 SBMOneK 0
    RW 0.4 0.9747 10 SBMOneK 0
    RW 0.5 0.8893 10 SBMOneK 0
    RW 0.6 0.3804 10 SBMOneK 0
    RW-1 0.05 1 10 SBMOneK 1
    RW-1 0.1 1 10 SBMOneK 1
    RW-1 0.2 1 10 SBMOneK 1
    RW-1 0.4 0.9699 10 SBMOneK 1
    RW-1 0.5 0.8962 10 SBMOneK 1
    RW-1 0.6 0.3826 10 SBMOneK 1
    RW-2 0.05 1 10 SBMOneK 2
    RW-2 0.1 1 10 SBMOneK 2
    RW-2 0.2 1 10 SBMOneK 2
    RW-2 0.4 0.9699 10 SBMOneK 2
    RW-2 0.5 0.8886 10 SBMOneK 2
    RW-2 0.6 0.4269 10 SBMOneK 2
    RW-3 0.05 1 10 SBMOneK 3
    RW-3 0.1 1 10 SBMOneK 3
    RW-3 0.2 1 10 SBMOneK 3
    RW-3 0.4 0.9728 10 SBMOneK 3
    RW-3 0.5 0.8959 10 SBMOneK 3
    RW-3 0.6 0.4031 10 SBMOneK 3
    RW-4 0.05 1 10 SBMOneK 4
    RW-4 0.1 1 10 SBMOneK 4
    RW-4 0.2 1 10 SBMOneK 4
    RW-4 0.4 0.9699 10 SBMOneK 4
    RW-4 0.5 0.8986 10 SBMOneK 4
    RW-4 0.6 0.4148 10 SBMOneK 4
    RW-5 0.05 1 10 SBMOneK 5
    RW-5 0.1 1 10 SBMOneK 5
    RW-5 0.2 1 10 SBMOneK 5
    RW-5 0.4 0.9728 10 SBMOneK 5
    RW-5 0.5 0.891 10 SBMOneK 5
    RW-5 0.6 0.3887 10 SBMOneK 5
    RW-10 0.05 1 10 SBMOneK 10
    RW-10 0.1 1 10 SBMOneK 10
    RW-10 0.2 1 10 SBMOneK 10
    RW-10 0.4 0.9699 10 SBMOneK 10
    RW-10 0.5 0.8943 10 SBMOneK 10
    RW-10 0.6 0.4212 10 SBMOneK 10
    RW-20 0.05 1 10 SBMOneK 20
    RW-20 0.1 1 10 SBMOneK 20
    RW-20 0.2 1 10 SBMOneK 20
    RW-20 0.4 0.9718 10 SBMOneK 20
    RW-20 0.5 0.8931 10 SBMOneK 20
    RW-20 0.6 0.3951 10 SBMOneK 20
    RW-50 0.05 1 10 SBMOneK 50
    RW-50 0.1 1 10 SBMOneK 50
    RW-50 0.2 1 10 SBMOneK 50
    RW-50 0.4 0.9699 10 SBMOneK 50
    RW-50 0.5 0.8893 10 SBMOneK 50
    RW-50 0.6 0.4635 10 SBMOneK 50
    RW-100 0.05 1 10 SBMOneK 100
    RW-100 0.1 1 10 SBMOneK 100
    RW-100 0.2 1 10 SBMOneK 100
    RW-100 0.4 0.9699 10 SBMOneK 100
    RW-100 0.5 0.8814 10 SBMOneK 100
    RW-100 0.6 0.4 10 SBMOneK 100
    RW-200 0.05 1 10 SBMOneK 200
    RW-200 0.1 1 10 SBMOneK 200
    RW-200 0.2 1 10 SBMOneK 200
    RW-200 0.4 0.9699 10 SBMOneK 200
    RW-200 0.5 0.8752 10 SBMOneK 200
    RW-200 0.6 0.3946 10 SBMOneK 200
    RW-500 0.05 1 10 SBMOneK 500
    RW-500 0.1 1 10 SBMOneK 500
    RW-500 0.2 1 10 SBMOneK 500
    RW-500 0.4 0.9671 10 SBMOneK 500
    RW-500 0.5 0.8888 10 SBMOneK 500
    RW-500 0.6 0.3893 10 SBMOneK 500
    RW-750 0.05 1 10 SBMOneK 750
    RW-750 0.1 1 10 SBMOneK 750
    RW-750 0.2 1 10 SBMOneK 750
    RW-750 0.4 0.9718 10 SBMOneK 750
    RW-750 0.5 0.8946 10 SBMOneK 750
    RW-750 0.6 0.3928 10 SBMOneK 750
    RW-1000 0.05 1 10 SBMOneK 1000
    RW-1000 0.1 1 10 SBMOneK 1000
    RW-1000 0.2 1 10 SBMOneK 1000
    RW-1000 0.4 0.9718 10 SBMOneK 1000
    RW-1000 0.5 0.8842 10 SBMOneK 1000
    RW-1000 0.6 0.3964 10 SBMOneK 1000
    RW-2000 0.05 1 10 SBMOneK 2000
    RW-2000 0.1 1 10 SBMOneK 2000
    RW-2000 0.2 1 10 SBMOneK 2000
    RW-2000 0.4 0.972 10 SBMOneK 2000
    RW-2000 0.5 0.8869 10 SBMOneK 2000
    RW-2000 0.6 0.4431 10 SBMOneK 2000
    RW-5000 0.05 1 10 SBMOneK 5000
    RW-5000 0.1 1 10 SBMOneK 5000
    RW-5000 0.2 1 10 SBMOneK 5000
    RW-5000 0.4 0.9699 10 SBMOneK 5000
    RW-5000 0.5 0.8947 10 SBMOneK 5000
    RW-5000 0.6 0.4189 10 SBMOneK 5000
}{\SBMOneKkTENallmethodsEC}

%%%%%%%%%%%%%%%%%

\pgfplotstableread[ignore chars={"}]{
    method mu EC avg_k family inertia
    Markov-Diff 0.05 1 15 SBMOneK 0
    Markov-Diff 0.1 1 15 SBMOneK 0
    Markov-Diff 0.2 1 15 SBMOneK 0
    Markov-Diff 0.4 1 15 SBMOneK 0
    Markov-Diff 0.5 0.9814 15 SBMOneK 0
    Markov-Diff 0.6 0.891 15 SBMOneK 0
    Markov-RW 0.05 1 15 SBMOneK 0
    Markov-RW 0.1 1 15 SBMOneK 0
    Markov-RW 0.2 1 15 SBMOneK 0
    Markov-RW 0.4 1 15 SBMOneK 0
    Markov-RW 0.5 0.9863 15 SBMOneK 0
    Markov-RW 0.6 0.8929 15 SBMOneK 0
    MultiLouvain 0.05 1 15 SBMOneK 0
    MultiLouvain 0.1 1 15 SBMOneK 0
    MultiLouvain 0.2 1 15 SBMOneK 0
    MultiLouvain 0.4 1 15 SBMOneK 0
    MultiLouvain 0.5 0.9853 15 SBMOneK 0
    MultiLouvain 0.6 0.8583 15 SBMOneK 0
    RW 0.05 1 15 SBMOneK 0
    RW 0.1 1 15 SBMOneK 0
    RW 0.2 1 15 SBMOneK 0
    RW 0.4 1 15 SBMOneK 0
    RW 0.5 0.9795 15 SBMOneK 0
    RW 0.6 0.8515 15 SBMOneK 0
    RW-1 0.05 1 15 SBMOneK 1
    RW-1 0.1 1 15 SBMOneK 1
    RW-1 0.2 1 15 SBMOneK 1
    RW-1 0.4 1 15 SBMOneK 1
    RW-1 0.5 0.9844 15 SBMOneK 1
    RW-1 0.6 0.8966 15 SBMOneK 1
    RW-2 0.05 1 15 SBMOneK 2
    RW-2 0.1 1 15 SBMOneK 2
    RW-2 0.2 1 15 SBMOneK 2
    RW-2 0.4 1 15 SBMOneK 2
    RW-2 0.5 0.9824 15 SBMOneK 2
    RW-2 0.6 0.8928 15 SBMOneK 2
    RW-3 0.05 1 15 SBMOneK 3
    RW-3 0.1 1 15 SBMOneK 3
    RW-3 0.2 1 15 SBMOneK 3
    RW-3 0.4 1 15 SBMOneK 3
    RW-3 0.5 0.9795 15 SBMOneK 3
    RW-3 0.6 0.8887 15 SBMOneK 3
    RW-4 0.05 1 15 SBMOneK 4
    RW-4 0.1 1 15 SBMOneK 4
    RW-4 0.2 1 15 SBMOneK 4
    RW-4 0.4 1 15 SBMOneK 4
    RW-4 0.5 0.9804 15 SBMOneK 4
    RW-4 0.6 0.8943 15 SBMOneK 4
    RW-5 0.05 1 15 SBMOneK 5
    RW-5 0.1 1 15 SBMOneK 5
    RW-5 0.2 1 15 SBMOneK 5
    RW-5 0.4 1 15 SBMOneK 5
    RW-5 0.5 0.9815 15 SBMOneK 5
    RW-5 0.6 0.8901 15 SBMOneK 5
    RW-10 0.05 1 15 SBMOneK 10
    RW-10 0.1 1 15 SBMOneK 10
    RW-10 0.2 1 15 SBMOneK 10
    RW-10 0.4 1 15 SBMOneK 10
    RW-10 0.5 0.9795 15 SBMOneK 10
    RW-10 0.6 0.8903 15 SBMOneK 10
    RW-20 0.05 1 15 SBMOneK 20
    RW-20 0.1 1 15 SBMOneK 20
    RW-20 0.2 1 15 SBMOneK 20
    RW-20 0.4 1 15 SBMOneK 20
    RW-20 0.5 0.9795 15 SBMOneK 20
    RW-20 0.6 0.8927 15 SBMOneK 20
    RW-50 0.05 1 15 SBMOneK 50
    RW-50 0.1 1 15 SBMOneK 50
    RW-50 0.2 1 15 SBMOneK 50
    RW-50 0.4 1 15 SBMOneK 50
    RW-50 0.5 0.9795 15 SBMOneK 50
    RW-50 0.6 0.8914 15 SBMOneK 50
    RW-100 0.05 1 15 SBMOneK 100
    RW-100 0.1 1 15 SBMOneK 100
    RW-100 0.2 1 15 SBMOneK 100
    RW-100 0.4 1 15 SBMOneK 100
    RW-100 0.5 0.9873 15 SBMOneK 100
    RW-100 0.6 0.8917 15 SBMOneK 100
    RW-200 0.05 1 15 SBMOneK 200
    RW-200 0.1 1 15 SBMOneK 200
    RW-200 0.2 1 15 SBMOneK 200
    RW-200 0.4 1 15 SBMOneK 200
    RW-200 0.5 0.9795 15 SBMOneK 200
    RW-200 0.6 0.8961 15 SBMOneK 200
    RW-500 0.05 1 15 SBMOneK 500
    RW-500 0.1 1 15 SBMOneK 500
    RW-500 0.2 1 15 SBMOneK 500
    RW-500 0.4 1 15 SBMOneK 500
    RW-500 0.5 0.9815 15 SBMOneK 500
    RW-500 0.6 0.891 15 SBMOneK 500
    RW-750 0.05 1 15 SBMOneK 750
    RW-750 0.1 1 15 SBMOneK 750
    RW-750 0.2 1 15 SBMOneK 750
    RW-750 0.4 1 15 SBMOneK 750
    RW-750 0.5 0.9824 15 SBMOneK 750
    RW-750 0.6 0.8877 15 SBMOneK 750
    RW-1000 0.05 1 15 SBMOneK 1000
    RW-1000 0.1 1 15 SBMOneK 1000
    RW-1000 0.2 1 15 SBMOneK 1000
    RW-1000 0.4 1 15 SBMOneK 1000
    RW-1000 0.5 0.9804 15 SBMOneK 1000
    RW-1000 0.6 0.8925 15 SBMOneK 1000
    RW-2000 0.05 1 15 SBMOneK 2000
    RW-2000 0.1 1 15 SBMOneK 2000
    RW-2000 0.2 1 15 SBMOneK 2000
    RW-2000 0.4 1 15 SBMOneK 2000
    RW-2000 0.5 0.9805 15 SBMOneK 2000
    RW-2000 0.6 0.8832 15 SBMOneK 2000
    RW-5000 0.05 1 15 SBMOneK 5000
    RW-5000 0.1 1 15 SBMOneK 5000
    RW-5000 0.2 1 15 SBMOneK 5000
    RW-5000 0.4 1 15 SBMOneK 5000
    RW-5000 0.5 0.9815 15 SBMOneK 5000
    RW-5000 0.6 0.8961 15 SBMOneK 5000
}{\SBMOneKkFIFTEENallmethodsEC}

%%%%%%%%%%%%%%%%%

\pgfplotstableread[ignore chars={"}]{
    method mu EC avg_k family inertia
    Markov-Diff 0.05 1 20 SBMOneK 0
    Markov-Diff 0.1 1 20 SBMOneK 0
    Markov-Diff 0.2 1 20 SBMOneK 0
    Markov-Diff 0.4 1 20 SBMOneK 0
    Markov-Diff 0.5 1 20 SBMOneK 0
    Markov-Diff 0.6 0.9834 20 SBMOneK 0
    Markov-RW 0.05 1 20 SBMOneK 0
    Markov-RW 0.1 1 20 SBMOneK 0
    Markov-RW 0.2 1 20 SBMOneK 0
    Markov-RW 0.4 1 20 SBMOneK 0
    Markov-RW 0.5 1 20 SBMOneK 0
    Markov-RW 0.6 0.9786 20 SBMOneK 0
    MultiLouvain 0.05 1 20 SBMOneK 0
    MultiLouvain 0.1 1 20 SBMOneK 0
    MultiLouvain 0.2 1 20 SBMOneK 0
    MultiLouvain 0.4 1 20 SBMOneK 0
    MultiLouvain 0.5 1 20 SBMOneK 0
    MultiLouvain 0.6 0.9892 20 SBMOneK 0
    RW 0.05 1 20 SBMOneK 0
    RW 0.1 1 20 SBMOneK 0
    RW 0.2 1 20 SBMOneK 0
    RW 0.4 1 20 SBMOneK 0
    RW 0.5 0.9971 20 SBMOneK 0
    RW 0.6 0.9795 20 SBMOneK 0
    RW-1 0.05 1 20 SBMOneK 1
    RW-1 0.1 1 20 SBMOneK 1
    RW-1 0.2 1 20 SBMOneK 1
    RW-1 0.4 1 20 SBMOneK 1
    RW-1 0.5 0.9971 20 SBMOneK 1
    RW-1 0.6 0.9805 20 SBMOneK 1
    RW-2 0.05 1 20 SBMOneK 2
    RW-2 0.1 1 20 SBMOneK 2
    RW-2 0.2 1 20 SBMOneK 2
    RW-2 0.4 1 20 SBMOneK 2
    RW-2 0.5 0.9971 20 SBMOneK 2
    RW-2 0.6 0.9776 20 SBMOneK 2
    RW-3 0.05 1 20 SBMOneK 3
    RW-3 0.1 1 20 SBMOneK 3
    RW-3 0.2 1 20 SBMOneK 3
    RW-3 0.4 1 20 SBMOneK 3
    RW-3 0.5 0.9971 20 SBMOneK 3
    RW-3 0.6 0.9776 20 SBMOneK 3
    RW-4 0.05 1 20 SBMOneK 4
    RW-4 0.1 1 20 SBMOneK 4
    RW-4 0.2 1 20 SBMOneK 4
    RW-4 0.4 1 20 SBMOneK 4
    RW-4 0.5 0.9971 20 SBMOneK 4
    RW-4 0.6 0.9776 20 SBMOneK 4
    RW-5 0.05 1 20 SBMOneK 5
    RW-5 0.1 1 20 SBMOneK 5
    RW-5 0.2 1 20 SBMOneK 5
    RW-5 0.4 1 20 SBMOneK 5
    RW-5 0.5 0.9971 20 SBMOneK 5
    RW-5 0.6 0.9776 20 SBMOneK 5
    RW-10 0.05 1 20 SBMOneK 10
    RW-10 0.1 1 20 SBMOneK 10
    RW-10 0.2 1 20 SBMOneK 10
    RW-10 0.4 1 20 SBMOneK 10
    RW-10 0.5 0.9971 20 SBMOneK 10
    RW-10 0.6 0.9796 20 SBMOneK 10
    RW-20 0.05 1 20 SBMOneK 20
    RW-20 0.1 1 20 SBMOneK 20
    RW-20 0.2 1 20 SBMOneK 20
    RW-20 0.4 1 20 SBMOneK 20
    RW-20 0.5 1 20 SBMOneK 20
    RW-20 0.6 0.9805 20 SBMOneK 20
    RW-50 0.05 1 20 SBMOneK 50
    RW-50 0.1 1 20 SBMOneK 50
    RW-50 0.2 1 20 SBMOneK 50
    RW-50 0.4 1 20 SBMOneK 50
    RW-50 0.5 0.9971 20 SBMOneK 50
    RW-50 0.6 0.9776 20 SBMOneK 50
    RW-100 0.05 1 20 SBMOneK 100
    RW-100 0.1 1 20 SBMOneK 100
    RW-100 0.2 1 20 SBMOneK 100
    RW-100 0.4 1 20 SBMOneK 100
    RW-100 0.5 0.9971 20 SBMOneK 100
    RW-100 0.6 0.9775 20 SBMOneK 100
    RW-200 0.05 1 20 SBMOneK 200
    RW-200 0.1 1 20 SBMOneK 200
    RW-200 0.2 1 20 SBMOneK 200
    RW-200 0.4 1 20 SBMOneK 200
    RW-200 0.5 0.9971 20 SBMOneK 200
    RW-200 0.6 0.9748 20 SBMOneK 200
    RW-500 0.05 1 20 SBMOneK 500
    RW-500 0.1 1 20 SBMOneK 500
    RW-500 0.2 1 20 SBMOneK 500
    RW-500 0.4 1 20 SBMOneK 500
    RW-500 0.5 0.9971 20 SBMOneK 500
    RW-500 0.6 0.9767 20 SBMOneK 500
    RW-750 0.05 1 20 SBMOneK 750
    RW-750 0.1 1 20 SBMOneK 750
    RW-750 0.2 1 20 SBMOneK 750
    RW-750 0.4 1 20 SBMOneK 750
    RW-750 0.5 0.998 20 SBMOneK 750
    RW-750 0.6 0.9805 20 SBMOneK 750
    RW-1000 0.05 1 20 SBMOneK 1000
    RW-1000 0.1 1 20 SBMOneK 1000
    RW-1000 0.2 1 20 SBMOneK 1000
    RW-1000 0.4 1 20 SBMOneK 1000
    RW-1000 0.5 0.9971 20 SBMOneK 1000
    RW-1000 0.6 0.9834 20 SBMOneK 1000
    RW-2000 0.05 1 20 SBMOneK 2000
    RW-2000 0.1 1 20 SBMOneK 2000
    RW-2000 0.2 1 20 SBMOneK 2000
    RW-2000 0.4 1 20 SBMOneK 2000
    RW-2000 0.5 0.9971 20 SBMOneK 2000
    RW-2000 0.6 0.9776 20 SBMOneK 2000
    RW-5000 0.05 1 20 SBMOneK 5000
    RW-5000 0.1 1 20 SBMOneK 5000
    RW-5000 0.2 1 20 SBMOneK 5000
    RW-5000 0.4 1 20 SBMOneK 5000
    RW-5000 0.5 0.9971 20 SBMOneK 5000
    RW-5000 0.6 0.9776 20 SBMOneK 5000
}{\SBMOneKkTWENTYallmethodsEC}

%%%%%%%%%%%%%%%%%

%% file: new-figs/cliq-ring-inertia.tex
\tikzstyle{clusterpair} = [
    draw=Orange, ellipse, inner sep=-1pt, dashed, %rotate fit=45,
    fill=Orange, fill opacity=0.20,
]
\tikzstyle{clustersingle} = [
    draw=NavyBlue, ellipse, inner sep=-0.125pt, densely dotted,
    fill=NavyBlue, fill opacity=0.25,
]

\pgfdeclarelayer{background layer}
\pgfdeclarelayer{foreground layer}
\pgfsetlayers{background layer,main,foreground layer}

\begin{tikzpicture}
  \begin{scope}[
    scale=3, 
    every node/.style={circle, draw, inner sep=0.6pt, fill=Gray!10, outer sep=0.5pt, font=\fontsize{2pt}{2pt}\selectfont},
    rotate=0,
    ]
      \def\n{30} % number of nodes
      \def\radius{0.85} % radius of the circle
    
      % Draw nodes in a circular layout
      \foreach \i in {1,...,\n} {
        % \node (\i) at ({\radius*cos(360/\n*\i)}, {\radius*sin(360/\n*\i)}) {K\textsubscript{5} \i};
        \node (\i) at ({\radius*sin(360/\n*\i)}, {\radius*cos(360/\n*\i)}) {K\textsubscript{5}};
      }
    
      % Draw edges between consecutive nodes to form a cycle
      \foreach \i [evaluate=\i as \next using {int(mod(\i, \n) + 1)}] in {1,...,\n} {
        \draw[gray, thick, draw opacity=0.5] (\i) -- (\next);
      }
    \end{scope}
    \begin{pgfonlayer}{background layer}      
        \node[clusterpair, rotate fit=0, fit=(1) (2)] {};
        \node[clusterpair, rotate fit=0, fit=(3) (4)] {};
        \node[clusterpair, rotate fit=0, fit=(5) (6)] {};
        \node[clusterpair, rotate fit=50, fit=(7) (8)] {};
        \node[clusterpair, fit=(9) (10)] {};
        \node[clusterpair, fit=(11) (12)] {};
        
        \node[clustersingle, fit=(13)] {};
        \node[clusterpair, rotate fit=60, fit=(14) (15)] {};
        \node[clustersingle, fit=(16)] {};
        
        \node[clusterpair, fit=(17) (18)] {};
        \node[clusterpair, fit=(19) (20)] {};
        
        \node[clustersingle, fit=(21)] {};
        
        \node[clusterpair, rotate fit=60, fit=(22) (23)] {};
        \node[clustersingle, fit=(24)] {};
        
        \node[clusterpair, rotate fit=20, fit=(25) (26)] {};
        \node[clusterpair, fit=(27) (28)] {};
        \node[clusterpair, rotate fit=-40, fit=(29) (30)] {};
    \end{pgfonlayer}
\end{tikzpicture}

%% file: new-figs/flights-world-sil-bar.tex
\input{data/flights-silhouette}

\newcommand\symcats{} % will hold "Alpha,Beta,Gamma"
\newif\iffirstcat
\firstcattrue  

\pgfplotstablegetrowsof{\FlightsWorldSilhouette}
\pgfmathtruncatemacro{\rows}{\pgfplotsretval-1}
\pgfplotsinvokeforeach{0,...,\rows}{%
  \pgfplotstablegetelem{#1}{method}\of{\FlightsWorldSilhouette}%
  \edef\thiscat{\pgfplotsretval}%
  %--- append, taking care of the initial empty case:
  \iffirstcat
      \xdef\symcats{\thiscat}%  first entry, no comma
      \firstcatfalse
  \else
      \xdef\symcats{\symcats,\thiscat}% prepend a comma thereafter
  \fi
}

\newcommand\symcatswt{} % will hold "Alpha,Beta,Gamma"
\newif\iffirstcat
\firstcattrue  

\pgfplotstablegetrowsof{\FlightsWorldSilhouetteWt}
\pgfmathtruncatemacro{\rows}{\pgfplotsretval-1}
\pgfplotsinvokeforeach{0,...,\rows}{%
  \pgfplotstablegetelem{#1}{method}\of{\FlightsWorldSilhouetteWt}%
  \edef\thiscat{\pgfplotsretval}%
  %--- append, taking care of the initial empty case:
  \iffirstcat
      \xdef\symcatswt{\thiscat}%  first entry, no comma
      \firstcatfalse
  \else
      \xdef\symcatswt{\symcatswt,\thiscat}% prepend a comma thereafter
  \fi
}

\begin{tikzpicture}
    \begin{groupplot}[
        group style={
            group size= 2 by 1,          % 2 columns × 1 row
            horizontal sep=75pt,
            ylabels at = edge left,      % first axis owns the y-axis labels
            % yticklabels at = edge left,
        },
        width=0.45\textwidth,                    % individual axis size
        height=9cm,
        xbar,
        /pgf/bar width=6pt,
        xmin=0, xmax=0.18,
        ytick style={draw=none},          % hide tick lines
        y dir   = reverse,                % keep first category at top
        scaled x ticks=true,
        xticklabel = {\pgfmathprintnumber[fixed,precision=2,zerofill]{\tick}},
        axis x line= bottom,
        axis y line*=middle,
        enlarge y limits=0.05,
        xlabel style={font=\small},
        xmajorgrids,
        unbounded coords=discard,	
        nodes near coords,
    ]
    
    % -------- First (left-hand) subplot -----------------
    \nextgroupplot[
        symbolic y coords/.expanded={\symcats},
        xmax=0.15,
        ytick=data,
        title={Mean Silhouette Score},
        every node near coord/.style={
            text=black,
            font=\fontsize{2pt}{2.5pt}\sffamily\selectfont,
            anchor=west,
            xshift=0pt,
            fill=white,
            /pgf/number format/.cd,
            fixed,          % fixed-point (no scientific “×10^k”)
            precision=3,    % three digits after the decimal
            zerofill,        % pad with trailing zeros (e.g. 0.500)
        },
    ]
        
        \addplot[
            draw=Turquoise,
            fill=SkyBlue!25,
        ]
        table[
            x=sil, y=method, meta=sil,
        ]{\FlightsWorldSilhouette};

    \nextgroupplot[
        symbolic y coords/.expanded={\symcatswt},
        ytick=data,
        xmax=0.15,
        title={Mean \textbf{Weighted} Silhouette Score},
        every node near coord/.style={
            text=black,
            font=\fontsize{2pt}{2.5pt}\sffamily\selectfont,
            fill=white,
            anchor=west,
            xshift=0pt,
            /pgf/number format/.cd,
            fixed,          % fixed-point (no scientific “×10^k”)
            precision=3,    % three digits after the decimal
            zerofill,        % pad with trailing zeros (e.g. 0.500)
        },
    ]
        \addplot[
            draw=Red,
            fill=Rhodamine!25,
        ]
        table[
            x=weighted_sil, y=method, meta=weighted_sil,
        ]{\FlightsWorldSilhouetteWt};

    \end{groupplot}
\end{tikzpicture}